\documentclass{emulateapj}

\usepackage{natbib}
\usepackage{amsmath}
\usepackage{multirow}

\newcommand{\ubvri}{\protect\hbox{$U\!BV\!RI$} }

\newcommand{\bvri}{\protect\hbox{$BV\!RI$} }

\newcommand{\about}{$\sim\!\!$~}
\newcommand{\kms}{\,km\,s$^{-1}$}

\def\lsim{\hbox{\rlap{\raise 0.425ex\hbox{$<$}}\lower 0.65ex\hbox{$\sim$}}}
\def\gsim{\hbox{\rlap{\raise 0.425ex\hbox{$>$}}\lower 0.65ex\hbox{$\sim$}}}
\def\arcmin{\hbox{$^\prime$}}

\def\arcdeg{\mbox{$^\circ$}}

\shorttitle{Early Observations of SN~2009ig}
\shortauthors{Foley et~al.}

\begin{document}

 \title{Very Early Ultraviolet and Optical Observations of the Type Ia
 Supernova 2009\lowercase{ig}}

\def\cfa{1}
\def\clay{2}
\def\berk{3}
\def\god{4}
\def\virg{5}
\def\ucsb{6}
\def\lbnl{7}
\def\berkphys{8}
\def\ucsc{9}
\def\stew{10}
\def\stsci{11}
\def\ifa{12}
\def\uci{13}
\def\hung{14}
\def\tex{15}
\def\seo{16}

\author{
{Ryan~J.~Foley}\altaffilmark{\cfa,\clay},
{P.~J.~Challis}\altaffilmark{\cfa},
{A.~V.~Filippenko}\altaffilmark{\berk},
{M.~Ganeshalingam}\altaffilmark{\berk},
{W.~Landsman}\altaffilmark{\god},
{W.~Li}\altaffilmark{\berk},
{G.~H.~Marion}\altaffilmark{\cfa},
{J.~M.~Silverman}\altaffilmark{\berk},
{R.~L.~Beaton}\altaffilmark{\virg},
{V.~N.~Bennert}\altaffilmark{\ucsb},
{S.~B.~Cenko}\altaffilmark{\berk},
{M.~Childress}\altaffilmark{\lbnl,\berkphys},
{P.~Guhathakurta}\altaffilmark{\ucsc},
{L.~Jiang}\altaffilmark{\stew},
{J.~S.~Kalirai}\altaffilmark{\stsci},
{R.~P.~Kirshner}\altaffilmark{\cfa},
{A.~Stockton}\altaffilmark{\ifa},
{E.~J.~Tollerud}\altaffilmark{\uci},
{J.~Vink\'{o}}\altaffilmark{\hung,\tex},
{J.~C.~Wheeler}\altaffilmark{\tex}, and
{J.-H.~Woo}\altaffilmark{\seo}
}

\altaffiltext{\cfa}{
Harvard-Smithsonian Center for Astrophysics,
60 Garden Street, 
Cambridge, MA 02138,
USA.
}
\altaffiltext{\clay}{
Clay Fellow. Electronic address rfoley@cfa.harvard.edu .
}
\altaffiltext{\berk}{
Department of Astronomy,
University of California,
Berkeley, CA 94720-3411,
USA.
}
\altaffiltext{\god}{
NASA Goddard Space Flight Center,
Greenbelt, MD 20771,
USA.
}
\altaffiltext{\virg}{
Department of Astronomy,
University of Virginia,
Charlottesville, VA 22904-4325,
USA.
}
\altaffiltext{\ucsb}{
Department of Physics,
University of California,
Santa Barbara, CA 93106-9530,
USA.
}
\altaffiltext{\lbnl}{
Physics Division,
Lawrence Berkeley National Laboratory,
1 Cyclotron Road,
Berkeley, CA 94720,
USA.
}
\altaffiltext{\berkphys}{
Department of Physics,
University of California,
366 LeConte Hall,
Berkeley, CA 94720-7300,
USA.
}
\altaffiltext{\ucsc}{
UCO/Lick Observatory,
University of California,
Santa Cruz, CA 95064,
USA.
}
\altaffiltext{\stew}{
Steward Observatory,
University of Arizona,
933 North Cherry Avenue,
Tucson, AZ 85721,
USA.
}
\altaffiltext{\stsci}{
Space Telescope Science Institute,
3700 San Martin Drive,
Baltimore, MD 21218,
USA.
}
\altaffiltext{\ifa}{
Institute for Astronomy,
University of Hawaii,
Honolulu, HI 96822,
USA.
}
\altaffiltext{\uci}{
Center for Cosmology,
Department of Physics and Astronomy,
4129 Frederick Reines Hall,
University of California,
Irvine, CA 92697,
USA.
}
\altaffiltext{\hung}{
Department of Optics \& Quantum Electronics,
University of Szeged,
D\'{o}m t\`{e}r 9, Szeged H-6720,
Hungary.
}
\altaffiltext{\tex}{
Astronomy Department,
University of Texas,
Austin, TX 78712,
USA.
}
\altaffiltext{\seo}{
Astronomy Program,
Department of Physics and Astronomy,
Seoul National University,
Seoul, 151-742,
Korea.
}

\begin{abstract}
Supernova (SN) 2009ig was discovered 17~hours after explosion by the
Lick Observatory Supernova Search, promptly classified as a normal
Type Ia SN (SN~Ia), peaked at $V = 13.5$~mag, and was equatorial,
making it one of the foremost supernovae for intensive study in the
last decade.  Here, we present ultraviolet (UV) and optical
observations of SN~2009ig, starting about 1~day after explosion until
around maximum brightness.  Our data include excellent UV and optical
light curves, 25 premaximum optical spectra, and 8 UV spectra,
including the earliest UV spectrum ever obtained of a SN~Ia.
SN~2009ig is a relatively normal SN~Ia, but does display high-velocity
ejecta --- the ejecta velocity measured in our earliest spectra ($v
\approx -23,000$~\kms\ for \ion{Si}{2} $\lambda 6355$) is the highest
yet measured in a SN~Ia.  The spectral evolution is very dramatic at
times earlier than 12~days before maximum brightness, but slows after
that time.  The early-time data provide a precise measurement of
$17.13 \pm 0.07$~days for the SN rise time.  The optical color curves
and early-time spectra are significantly different from template light
curves and spectra used for light-curve fitting and $K$-corrections,
indicating that the template light curves and spectra do not properly
represent all Type Ia supernovae at very early times.  In the age of
wide-angle sky surveys, SNe like SN~2009ig that are nearby, bright,
well positioned, and promptly discovered will still be rare.  As shown
with SN~2009ig, detailed studies of single events can provide
significantly more information for testing systematic uncertainties
related to SN~Ia distance estimates and constraining progenitor and
explosion models than large samples of more distant SNe.
\end{abstract}



\section{Introduction}\label{s:intro}

Type Ia supernovae (SNe~Ia) are exceptionally good distance
indicators, allowing precise measurements of various cosmological
parameters, including the first significant constraints on
$\Omega_{\Lambda}$ \citep[e.g.,][]{Riess98:lambda, Perlmutter99,
Riess07, Wood-Vasey07, Hicken09:de, Kessler09:cosmo, Amanullah10,
Conley11, Suzuki11}.  The progenitor system and explosion mechanism
are generally known \citep[a thermonuclear explosion of a C/O white
dwarf (WD) in a binary system;][]{Hoyle60, Colgate69, Nomoto84:w7,
Woosley86}, but the specifics of the various models --- such as
whether the progenitor comes from a single- or double-degenerate
system --- are ill constrained \citep[e.g.,][]{Hillebrandt00,
Howell11}.  Many studies of individual peculiar SNe~Ia, which can
probe the extremities of the models or show what a normal SN~Ia is
{\it not}, have been conducted over the years
\citep[e.g.,][]{Li01:00cx, Li03:02cx, Hamuy03, Howell06, Foley09:08ha,
Foley10:06bt}, but intense studies of individual normal SNe~Ia are
relatively rare.  Systematic studies of large samples of normal SNe~Ia
can provide important constraints for the models, but extremely
detailed observations of even a single normal object can be as
constraining as hundreds of objects with much lower quality data.

The ultraviolet (UV) portion of a SN~Ia spectral energy distribution
(SED) provides a bounty of information about the explosions of SNe~Ia.
UV spectra are the most constraining data for determining the effects
of temperature, density, and nonthermal ionization
\citep[e.g.,][]{Hoflich98, Lentz01}. SN~Ia UV spectra are dominated by
a forest of overlapping lines from Fe-group elements.  UV photons are
repeatedly absorbed and re-emitted in those lines and gradually
scattered redward where lower opacities allow them to escape.
Therefore, the UV is crucial to the formation of the optical SED of
SNe~Ia \citep{Sauer08} and extremely sensitive to both the progenitor
composition and explosion mechanism.  Observations of the UV directly
probe the composition of the outermost layers of ejecta which are
transparent at optical wavelengths soon after explosion.  Furthermore,
\citet{Kasen10:prog} showed that single-degenerate progenitor systems
should have strong UV emission at early times from the SN ejecta
interacting with the companion star; therefore, early UV observations
directly test progenitor models.

At optical wavelengths SNe~Ia are remarkably uniform in luminosity
($\sigma \approx 0.16$~mag; \citealt{Mandel11}), after correcting for
light-curve shape and color.  This relationship extends to the $U$
band and UV, but with larger scatter \citep{Jha06:lc, Brown10}.  The
scatter in the optical can be further reduced if one accounts for the
correlation between intrinsic color and ejecta velocity
\citep{Foley11:vel}.  Along with the observed $B-V$ color of SNe~Ia
\citep{Pignata08:02dj, Wang09:2pop}, their intrinsic $B-V$ color
correlates strongly with velocity, with intrinsically redder objects
having high velocity \citep{Foley11:vel, Foley11:vgrad}.  This is
explained as additional line blanketing in the $B$ band of the
high-velocity objects, and this trend should extend to the UV
\citep{Foley11:vel}.  \citet{Ganeshalingam11} found that
higher-velocity SNe~Ia tend to have faster $B$-band rise times than
lower-velocity SNe~Ia ($\Delta t = 1.4$~days), but this is not the case
for $V$-band rise times.

SNe~Ia bright enough to be observed by the {\it International
Ultraviolet Explorer} ({\it IUE}) were rare \citep{Foley08:uv}, and
only one object has a published high signal-to-noise ratio (S/N) {\it
Hubble Space Telescope} ({\it HST}) spectrum near maximum light
covering wavelengths $\lesssim 2900$~\AA: SN~1992A \citep{Kirshner93},
observed nearly two decades ago.  {\it Swift} has obtained spectra of
several SNe~Ia, but because of typically short exposure times, they
are generally of poor quality \citep{Bufano09}.  In Cycle 17, {\it
HST} obtained a single spectrum of 30 SNe~Ia (GO-11721; PI Ellis);
however, because of the choice of grating, these spectra do not probe
wavelengths shorter than \about 2900~\AA\ \citep{Cooke11}.  In total,
there have been 7 SNe~Ia with published premaximum spectra probing
wavelengths shortward of \about 2900~\AA\ (SNe~1980N, 1986G, 1989B,
1990N, 1991T, 1992A, and 2005cf; \citealt{Jeffery92, Kirshner93,
Foley08:uv, Bufano09}).

SN~2009ig was discovered at an unfiltered magnitude of 17.5, 4 mag
below peak, on 2009 August 20.48 (UT dates are used throughout this
paper) by \citet{Kleiser09} during the Lick Observatory Supernova
Search (LOSS) with the 0.76~m Katzman Automatic Imaging Telescope
\citep[KAIT;][]{Filippenko01}.  There was no object detected on 2009
August 16.46 to a limit of 18.7~mag.  It was discovered in NGC~1015,
an SBa galaxy at $cz = 2629$~\kms \citep{Wong06} with distance $D =
33.1$~Mpc ($\mu = 32.6 \pm 0.4$~mag) from a Tully-Fisher measurement
\citep{Tully88}.

\citet{Navasardyan09} obtained an optical spectrum of SN~2009ig on
2009 August 21.08, only 0.7~days after discovery, and determined that
it was a young SN~Ia.  Clearly, SN~2009ig was discovered very shortly
after explosion.  We triggered multiple programs to study the
photometric and spectroscopic evolution of the object, its
circumstellar environment, its spectropolarimetry, its energetics, and
other aspects.  Here we focus on the early-time UV and optical SN
spectroscopy and photometry.

We present our UV and optical data in Section~\ref{s:obs}, and derive
basic maximum-light parameters for SN~2009ig in Section~\ref{s:basic}.
The UV spectra are discussed in Section~\ref{s:uv}.  In
Section~\ref{s:ep} and Section~\ref{s:os}, we compare the early-time
photometry and spectroscopy, respectively, to that of other objects
and to models.  We discuss our findings and summarize our conclusions
in Section~\ref{s:disc}.


\section{Observations and Data Reduction}\label{s:obs}

\subsection{Optical and Ultraviolet Photometry}

Broad-band \bvri\ photometry of SN~2009ig was obtained using KAIT with
the KAIT4 filters set \citep{Ganeshalingam10} at Lick Observatory
starting on 21.5 August 2009 (about two weeks before maximum light).
Within a day of the discovery of SN~2009ig, KAIT was programmed to
observe the field with a nightly cadence to sample the rise and
maximum of the light curves.  Two weeks after maximum light the
cadence was changed to every 3--4~days.  In total, we have 55
photometry epochs.  A finder chart of SN~2009ig, its host galaxy, and
a comparison star is shown in Figure~\ref{f:finder}.

\begin{figure}
\begin{center}
\epsscale{1.1}
\rotatebox{0}{
\plotone{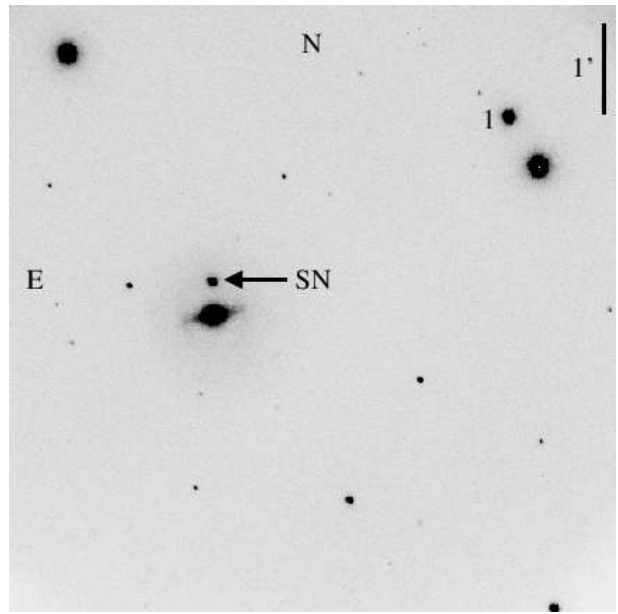}}
\caption{KAIT $R$-band image of SN~2009ig and its host galaxy,
NGC~1015.  The field of view (FOV) is 6.7\arcmin $\times$ 6.7\arcmin.
The SN and comparison star are marked.}\label{f:finder}
\end{center}
\end{figure}

The data were reduced using a mostly automated pipeline developed for
KAIT images \citep{Ganeshalingam10}.  Images were bias-corrected and
flatfielded at the telescope.  Using galaxy templates obtained a year
and a half after discovery, the data images were galaxy subtracted to
remove galaxy flux at the position of the SN.  The flux of the SN and
the local field star were measured using the point-spread function
(PSF) fitting photometry package \texttt{DAOPHOT} in
IRAF\footnote{IRAF: The Image Reduction and Analysis Facility is
distributed by the National Optical Astronomy Observatory, which is
operated by the Association of Universities for Research in Astronomy,
Inc., under cooperative agreement with the National Science Foundation
(NSF).}.  Instrumental magnitudes were color-corrected to the
\citet{Landolt92} system using the average color terms measured on
multiple photometric nights as presented by \citet{Ganeshalingam10}.
The magnitudes of the local field standard were calibrated against
\citet{Landolt92} standards on seven photometric nights.

The field of SN~2009ig suffers from a dearth of local standards that
are bright enough to measure reliably, but not so bright that they
saturate the detector.  After trying combinations of available
standards, we find that the smoothest, most believable light curves
are obtained using a single, bright comparison star whose photometry
is given in Table~\ref{t:standard}.  We caution that this leaves our
final light curves susceptible to a systematic offset from an
incorrect determination of the Landolt magnitude of the comparison
star, but should not change the overall shape of our light curves.
KAIT photometry of SN~2009ig is presented in Figure~\ref{f:lc} and
Table~\ref{t:kait}.

\begin{deluxetable*}{ccccccccccc}
\tablewidth{0pc}
\tablecaption{Photometry of SN~2009ig Standard Star\label{t:standard}}
\tablehead { \colhead{ID} & \colhead{$\alpha$ (J2000)} & \colhead{$\delta$ (J2000)} & \colhead{$B (\sigma_{B})$} & \colhead{$N_{B}$} & \colhead{$V (\sigma_{V})$} & \colhead{$N_{V}$} & \colhead{$R (\sigma_{R})$} & \colhead{$N_{R}$} & \colhead{$I (\sigma_{I})$} & \colhead{$N_{I}$}}

\startdata

 1 &  02:37:58.65 & $-01$:16:56.4 & 13.637  (004) &   6 & 13.039  (003)  & 6 &  12.829  (007) &   4 &  12.426  (015) &   3

\enddata

\tablecomments{Photometry given in magnitudes with 1$\sigma$
uncertainties (in mmags) presented in parentheses.  $N$ corresponds to
the number of calibration images for each band.}

\end{deluxetable*}

\begin{figure}
\begin{center}
\epsscale{2.}
\rotatebox{90}{
\plotone{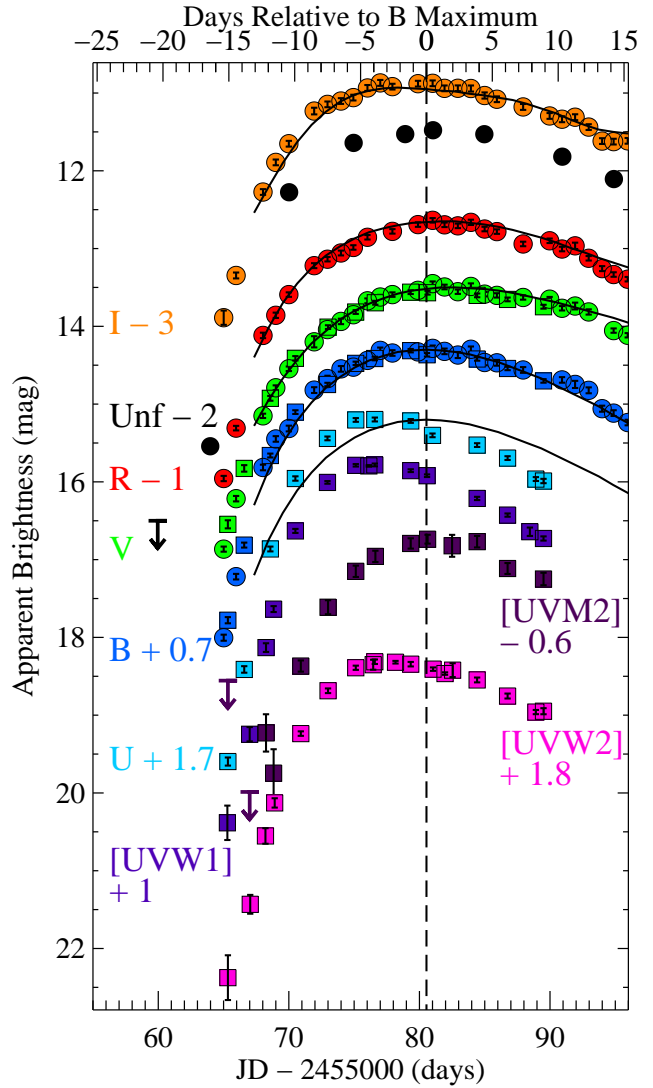}}
\caption{[$UVW2$][$UV\!M2$][$UVW1$]\ubvri\ (fuchsia, dark purple, navy,
cyan, blue, green, red, and orange, respectively; circles and squares
correspond to KAIT and {\it Swift}/UVOT data, respectively) and
unfiltered (black circles, with the label ``Unf'') light curves of
SN~2009ig.  Our unfiltered magnitudes closely approximate the $R$
band.  The uncertainties for most data points are smaller than the
plotted symbols. Also shown are comparison \ubvri\ template light
curves of SN~2005cf \citep{Wang09:05cf}, stretched by a factor of
1.1.}\label{f:lc}
\end{center}
\end{figure}

\begin{deluxetable*}{lcrcrcrcr}
\tablewidth{0pc}
\tablecaption{KAIT Photometry and S-Corrections for SN~2009ig\label{t:kait}}
\tablehead{\colhead{JD} & \colhead{$B$ (mag)} & \colhead{$S_{B}$ (mag)} & \colhead{$V$ (mag)} & \colhead{$S_{V}$ (mag)} & \colhead{$R$ (mag)} & \colhead{$S_{R}$ (mag)} & \colhead{$I$ (mag)} & \colhead{$S_{I}$ (mag)} }

\startdata

2455065.01 & 17.304 (039) & $-0.110$ & 16.865 (033) &   0.010  & 16.957 (034) &   0.020  & 16.889 (092) & $-0.020$ \\
2455065.96 & 16.519 (036) & $-0.100$ & 16.216 (035) &   0.000  & 16.309 (029) &   0.020  & 16.346 (038) & $-0.020$ \\
2455068.02 & 15.115 (036) & $-0.050$ & 15.151 (028) & $-0.010$ & 15.117 (035) &   0.010  & 15.273 (037) & $-0.010$ \\
2455069.00 & 14.748 (036) & $-0.040$ & 14.790 (031) & $-0.010$ & 14.858 (034) &   0.010  & 14.892 (037) & $-0.010$ \\
2455070.00 & 14.614 (037) & $-0.040$ & 14.546 (029) & $-0.010$ & 14.592 (029) &   0.010  & 14.652 (037) & $-0.010$ \\
2455071.94 & 14.120 (037) & $-0.030$ & 14.199 (072) & $-0.010$ & 14.220 (032) &   0.000  & 14.232 (043) & $-0.010$ \\
2455072.96 & 13.972 (037) & $-0.030$ & 14.042 (028) & $-0.010$ & 14.137 (029) &   0.010  & 14.146 (037) & $-0.010$ \\
2455073.99 & 13.847 (036) & $-0.030$ & 13.937 (035) & $-0.010$ & 14.057 (031) &   0.010  & 14.103 (037) & $-0.010$ \\
2455074.94 & 13.825 (036) & $-0.030$ & 13.851 (028) &   0.000  & 13.986 (029) &   0.010  & 14.059 (037) & $-0.010$ \\
2455076.02 & 13.738 (036) & $-0.020$ & 13.672 (031) &   0.000  & 13.855 (029) &   0.020  & 13.935 (038) & $-0.020$ \\
2455077.00 & 13.609 (036) & $-0.030$ & 13.623 (028) &   0.000  & \nodata      &   0.020  & 13.870 (038) & $-0.020$ \\
2455077.93 & 13.643 (036) & $-0.030$ & 13.592 (029) &   0.000  & 13.779 (029) & \nodata  & 13.918 (037) & $-0.020$ \\
2455079.89 & 13.629 (036) & $-0.040$ & 13.537 (028) &   0.000  & 13.692 (029) &   0.030  & 13.880 (037) & $-0.020$ \\
2455081.01 & 13.583 (036) & $-0.030$ & 13.451 (028) &   0.000  & 13.637 (029) &   0.030  & 13.875 (037) & $-0.030$ \\
2455081.92 & 13.622 (036) & $-0.030$ & 13.494 (030) &   0.000  & 13.695 (029) &   0.030  & 13.940 (037) & $-0.030$ \\
2455082.94 & 13.669 (039) & $-0.030$ & 13.555 (028) &   0.000  & 13.708 (029) &   0.030  & 13.939 (037) & $-0.030$ \\
2455083.94 & 13.595 (036) & $-0.030$ & 13.487 (039) &   0.000  & 13.663 (034) &   0.030  & 13.942 (039) & $-0.030$ \\
2455084.99 & 13.765 (036) & $-0.030$ & 13.586 (029) &   0.000  & 13.749 (029) &   0.040  & 14.032 (040) & $-0.030$ \\
2455085.93 & 13.769 (037) & $-0.030$ & 13.602 (028) &   0.000  & 13.782 (029) &   0.030  & 14.080 (037) & $-0.030$ \\
2455087.94 & 13.861 (037) & $-0.030$ & 13.629 (028) &   0.000  & 13.941 (029) &   0.030  & 14.179 (038) & $-0.040$ \\
2455089.97 & \nodata      & \nodata  & 13.651 (029) & $-0.010$ & 13.900 (029) &   0.030  & 14.295 (037) & $-0.040$ \\
2455090.92 & 13.989 (036) & $-0.030$ & 13.768 (028) & $-0.010$ & 14.006 (029) &   0.020  & 14.337 (038) & $-0.040$ \\
2455091.92 & 14.047 (046) & $-0.040$ & 13.739 (044) & $-0.010$ & 13.965 (041) &   0.020  & 14.311 (038) & $-0.050$ \\
2455092.96 & 14.123 (038) & $-0.040$ & 13.815 (034) &   0.000  & 14.125 (029) &   0.020  & 14.438 (037) & $-0.050$ \\
2455093.99 & 14.359 (039) & $-0.040$ & \nodata      & \nodata  & 14.258 (032) &   0.020  & 14.618 (037) & $-0.050$ \\
2455094.87 & 14.416 (037) & $-0.021$ & 14.055 (028) &   0.000  & 14.334 (029) &   0.010  & 14.625 (037) & $-0.050$ \\
2455095.95 & 14.537 (036) & $-0.020$ & 14.115 (028) &   0.000  & 14.394 (029) &   0.010  & 14.615 (037) & $-0.050$ \\
2455097.02 & 14.639 (038) & $-0.020$ & 14.085 (030) &   0.000  & 14.345 (029) &   0.010  & 14.606 (037) & $-0.050$ \\
2455097.90 & 14.776 (038) & $-0.020$ & 14.238 (028) &   0.000  & 14.460 (029) &   0.010  & 14.663 (037) & $-0.050$ \\
2455098.93 & 14.885 (036) & $-0.030$ & 14.203 (028) &   0.010  & 14.482 (029) &   0.010  & 14.621 (037) & $-0.060$ \\
2455099.94 & 15.000 (036) & $-0.020$ & 14.301 (029) &   0.010  & 14.470 (029) &   0.020  & 14.601 (037) & $-0.050$ \\
2455100.93 & 15.113 (042) & $-0.030$ & \nodata      & \nodata  & 14.444 (031) &   0.020  & 14.515 (037) & $-0.050$ \\
2455101.98 & 15.145 (037) & $-0.040$ & 14.330 (028) &   0.010  & 14.476 (029) &   0.020  & 14.461 (040) & $-0.050$ \\
2455102.96 & 15.175 (036) & $-0.050$ & 14.335 (028) &   0.010  & 14.425 (029) &   0.020  & 14.435 (037) & $-0.040$ \\
2455104.96 & 15.381 (036) & $-0.060$ & 14.413 (028) &   0.010  & 14.403 (029) &   0.030  & 14.394 (037) & $-0.030$ \\
2455106.87 & 15.591 (039) & $-0.050$ & 14.524 (029) &   0.010  & 14.425 (029) &   0.030  & 14.434 (037) & $-0.020$ \\
2455112.91 & 16.076 (044) & $-0.060$ & 14.825 (029) &   0.020  & 14.666 (030) &   0.030  & \nodata      & \nodata  \\
2455114.90 & 16.112 (037) & $-0.060$ & 14.928 (038) &   0.020  & 14.781 (029) &   0.020  & 14.430 (037) & $-0.010$ \\
2455116.96 & 16.206 (091) & $-0.060$ & 15.104 (029) &   0.010  & 14.916 (029) &   0.020  & 14.544 (037) & $-0.010$ \\
2455120.93 & 16.288 (036) & $-0.080$ & 15.292 (028) &   0.010  & 15.066 (031) &   0.020  & 14.754 (037) &  0.000   \\
2455123.89 & 16.480 (039) & $-0.080$ & 15.388 (028) &   0.010  & 15.337 (029) &   0.030  & 15.049 (037) & $-0.010$ \\
2455126.91 & 16.418 (036) & $-0.060$ & 15.448 (043) &   0.010  & 15.363 (030) &   0.030  & 15.169 (037) &  0.000   \\
2455129.81 & 16.514 (036) & $-0.050$ & 15.636 (029) &   0.010  & 15.532 (029) &   0.020  & 15.325 (037) &  0.000   \\
2455132.86 & 16.619 (047) & $-0.050$ & 15.721 (031) &   0.010  & 15.572 (029) &   0.020  & 15.442 (038) &  0.010   \\
2455133.85 & 16.546 (040) & $-0.040$ & 15.704 (029) &   0.000  & 15.544 (034) &   0.020  & 15.509 (037) &  0.020   \\
2455139.91 & 16.716 (063) & $-0.040$ & 15.941 (044) &   0.000  & 15.809 (088) &   0.020  & 15.880 (061) &  0.020   \\
2455143.81 & 17.001 (148) & $-0.050$ & 16.088 (043) &   0.000  & 15.935 (031) &   0.020  & 15.973 (047) &  0.010   \\
2455148.89 & 16.792 (044) & $-0.020$ & 16.081 (038) &   0.010  & 16.074 (029) &   0.010  & 16.063 (037) &  0.010   \\
2455154.82 & 16.974 (037) &   0.000  & 16.316 (029) &   0.010  & 16.326 (029) &   0.000  & 16.397 (038) &  0.010   \\
2455159.77 & 16.983 (036) &   0.020  & 16.376 (029) &   0.000  & 16.402 (029) &   0.000  & 16.490 (040) &  0.010   \\
2455168.81 & 17.187 (045) &   0.040  & 16.682 (033) &   0.000  & \nodata      & \nodata  & 16.820 (038) &  0.010   \\
2455173.83 & 17.295 (037) &   0.030  & 16.745 (030) &   0.000  & 16.835 (030) & $-0.010$ & 16.981 (040) &  0.010   \\
2455182.76 & 17.286 (036) &   0.060  & 16.994 (030) &   0.000  & 17.066 (041) & $-0.010$ & 17.385 (127) &  0.010   \\
2455188.71 & 17.527 (036) &   0.050  & 17.095 (033) &   0.000  & 17.263 (031) & $-0.010$ & 17.509 (069) &  0.010   \\
2455196.79 & 17.589 (061) &   0.060  & \nodata      & \nodata  & \nodata      & \nodata  & \nodata	& \nodata  

\enddata

\tablecomments{Photometry given in magnitudes with 1$\sigma$
(photometric and calibration) uncertainties (in mmags) presented in
parentheses.  The noted $S$-corrections have been applied to the
photometry.}

\end{deluxetable*}

The {\it Swift} team initiated target-of-opportunity observations of
SN~2009ig with the Ultraviolet/Optical Telescope (UVOT;
\citealt{Roming05}) and the X-ray Telescope (XRT; \citealt{Burrows05})
onboard the {\it Swift} gamma-ray burst satellite \citep{Gehrels04}
beginning 2009 August 21.8.  We performed digital image subtraction on
all of the UVOT data using the final epoch as a template to remove
host-galaxy contamination with the ISIS software package
\citep{Alard00}.  The $U$-, $B$-, and $V$-band data were then reduced
using the calibration technique described by \citet{Li06}, while
[$UVW2$], [$UV\!M2$], and [$UVW1$] (corresponding to central/effective
wavelengths of 1941/3064, 2248/2360, and 2605/3050~\AA, respectively;
\citealt{Brown10}) photometry was obtained using the zeropoints from
\citet{Poole08}.

The three bluest UVOT filters all have a significant wing that extends
to optical wavelengths.  This ``red leak'' is particularly problematic
for SNe~Ia, which have SEDs that peak in the optical and significantly
depressed UV flux relative to the optical flux.  As a result, a
significant fraction of the flux measured in the UV filters comes from
the optical portion of the SED.  \citet{Brown10} provides red-leak
corrections for the UVOT filters.  However, these corrections were
calculated from the $t = +5$~day SN~1992A spectrum, which is the
single published high S/N UV near-maximum SN~Ia spectrum
\citep{Kirshner93}.  SN~Ia SEDs evolve with time, and especially
quickly for the phases we examine here.  We therefore do not want to
make a single correction to our data.  Unfortunately, there are no
other suitable, published SN~Ia spectra from which we could make
red-leak corrections.  As a result, we present the UVOT data without a
red-leak correction.

The results of our UVOT analysis are displayed in Figure~\ref{f:lc}
and presented in Table~\ref{t:swift}.

\begin{deluxetable}{lclr}
\tablewidth{0pc}
\tablecaption{{\it Swift}/UVOT Photometry of SN~2009ig\label{t:swift}}
\tablehead{\colhead{JD} & \colhead{Filter} & \colhead{Magnitude} & \colhead{$S$-Correction (mag)}}

\startdata

2455065.33 & [$UVW2$] & 20.575 (288) & \nodata \\
2455067.04 & [$UVW2$] & 19.632 (121) & \nodata \\
2455068.21 & [$UVW2$] & 18.753 (100) & \nodata \\
2455068.92 & [$UVW2$] & 18.328 (060) & \nodata \\
2455070.92 & [$UVW2$] & 17.437 (029) & \nodata \\
2455073.00 & [$UVW2$] & 16.886 (030) & \nodata \\
2455075.13 & [$UVW2$] & 16.590 (027) & \nodata \\
2455076.46 & [$UVW2$] & 16.551 (064) & \nodata \\
2455076.58 & [$UVW2$] & 16.513 (029) & \nodata \\
2455078.17 & [$UVW2$] & 16.519 (017) & \nodata \\
2455079.33 & [$UVW2$] & 16.544 (029) & \nodata \\
2455081.04 & [$UVW2$] & 16.608 (022) & \nodata \\
2455081.96 & [$UVW2$] & 16.665 (016) & \nodata \\
2455082.54 & [$UVW2$] & 16.621 (089) & \nodata \\
2455084.42 & [$UVW2$] & 16.747 (031) & \nodata \\
2455086.75 & [$UVW2$] & 16.954 (032) & \nodata \\
2455088.92 & [$UVW2$] & 17.160 (027) & \nodata \\
2455089.50 & [$UVW2$] & 17.149 (039) & \nodata \\
2455098.96 & [$UVW2$] & 18.295 (071) & \nodata \\
2455105.96 & [$UVW2$] & 18.761 (089) & \nodata \\
2455107.88 & [$UVW2$] & 19.067 (113) & \nodata \\
2455113.79 & [$UVW2$] & 19.199 (112) & \nodata \\
2455116.88 & [$UVW2$] & 19.297 (122) & \nodata \\
2455065.33 & [$UV\!M2$] & $>19.156$ & \nodata \\
2455067.00 & [$UV\!M2$] & $>20.587$ & \nodata \\
2455068.25 & [$UV\!M2$] & 19.828 (240) & \nodata \\
2455068.83 & [$UV\!M2$] & 20.345 (307) & \nodata \\
2455070.92 & [$UV\!M2$] & 18.968 (072) & \nodata \\
2455073.00 & [$UV\!M2$] & 18.212 (090) & \nodata \\
2455075.13 & [$UV\!M2$] & 17.750 (073) & \nodata \\
2455076.63 & [$UV\!M2$] & 17.558 (073) & \nodata \\
2455079.33 & [$UV\!M2$] & 17.392 (066) & \nodata \\
2455080.63 & [$UV\!M2$] & 17.337 (057) & \nodata \\
2455082.50 & [$UV\!M2$] & 17.422 (141) & \nodata \\
2455084.42 & [$UV\!M2$] & 17.369 (074) & \nodata \\
2455086.75 & [$UV\!M2$] & 17.714 (070) & \nodata \\
2455089.50 & [$UV\!M2$] & 17.849 (080) & \nodata \\
2455098.96 & [$UV\!M2$] & 18.912 (147) & \nodata \\
2455105.96 & [$UV\!M2$] & 19.480 (203) & \nodata \\
2455107.88 & [$UV\!M2$] & 19.578 (220) & \nodata \\
2455113.79 & [$UV\!M2$] & 20.346 (326) & \nodata \\
2455116.88 & [$UV\!M2$] & $>20.100$ & \nodata \\
2455065.29 & [$UVW1$] & 19.384 (221) & \nodata \\
2455067.00 & [$UVW1$] & 18.245 (092) & \nodata \\
2455068.25 & [$UVW1$] & 17.131 (056) & \nodata \\
2455068.83 & [$UVW1$] & 16.636 (038) & \nodata \\
2455070.50 & [$UVW1$] & 15.631 (026) & \nodata \\
2455072.96 & [$UVW1$] & 15.007 (017) & \nodata \\
2455075.13 & [$UVW1$] & 14.787 (016) & \nodata \\
2455076.13 & [$UVW1$] & 14.795 (010) & \nodata \\
2455076.58 & [$UVW1$] & 14.782 (018) & \nodata \\
2455079.33 & [$UVW1$] & 14.856 (018) & \nodata \\
2455080.58 & [$UVW1$] & 14.921 (017) & \nodata \\
2455084.42 & [$UVW1$] & 15.213 (021) & \nodata \\
2455086.75 & [$UVW1$] & 15.427 (021) & \nodata \\
2455088.46 & [$UVW1$] & 15.641 (054) & \nodata \\
2455089.50 & [$UVW1$] & 15.728 (027) & \nodata \\
2455098.96 & [$UVW1$] & 16.846 (049) & \nodata \\
2455101.83 & [$UVW1$] & 17.145 (073) & \nodata \\
2455105.92 & [$UVW1$] & 17.609 (072) & \nodata \\
2455107.88 & [$UVW1$] & 17.713 (081) & \nodata \\
2455113.79 & [$UVW1$] & 18.035 (090) & \nodata \\
2455116.88 & [$UVW1$] & 18.481 (123) & \nodata

\enddata

\end{deluxetable}

\begin{deluxetable}{lclr}
\tablewidth{0pc}
\tablecaption{{\it Swift}/UVOT Photometry of SN~2009ig (continued)\label{t:swift2}}
\tablehead{\colhead{JD} & \colhead{Filter} & \colhead{Magnitude} & \colhead{$S$-Correction (mag)}}

\startdata

2455065.33 & $U$ & 17.896 (068) & \nodata  \\
2455066.58 & $U$ & 16.712 (042) & \nodata  \\
2455068.58 & $U$ & 15.163 (021) & \nodata  \\
2455070.50 & $U$ & 14.257 (015) & \nodata  \\
2455072.96 & $U$ & 13.741 (011) & \nodata  \\
2455075.13 & $U$ & 13.504 (011) & \nodata  \\
2455076.58 & $U$ & 13.498 (012) & \nodata  \\
2455079.33 & $U$ & 13.517 (012) & \nodata  \\
2455081.00 & $U$ & 13.702 (006) & \nodata  \\
2455084.42 & $U$ & 13.827 (012) & \nodata  \\
2455086.75 & $U$ & 13.994 (012) & \nodata  \\
2455088.92 & $U$ & 14.264 (005) & \nodata  \\
2455089.50 & $U$ & 14.289 (015) & \nodata  \\
2455098.96 & $U$ & 15.385 (024) & \nodata  \\

2455065.33 & $B$ & 17.079 (040) & $-0.047$ \\
2455066.58 & $B$ & 16.113 (032) & $-0.031$ \\
2455068.58 & $B$ & 14.961 (025) & $-0.022$ \\
2455070.50 & $B$ & 14.403 (024) & $-0.011$ \\
2455072.96 & $B$ & 14.048 (022) & $-0.012$ \\
2455075.12 & $B$ & 13.776 (022) & $-0.008$ \\
2455076.58 & $B$ & 13.714 (023) & $-0.002$ \\
2455079.33 & $B$ & 13.613 (022) & $-0.003$ \\
2455080.58 & $B$ & 13.666 (022) & $-0.005$ \\
2455084.42 & $B$ & 13.720 (022) & $-0.005$ \\
2455086.75 & $B$ & 13.842 (022) & $-0.008$ \\
2455089.50 & $B$ & 14.002 (023) & $-0.010$ \\
2455098.96 & $B$ & 14.845 (025) & $-0.011$ \\
2455105.96 & $B$ & 15.483 (028) & $-0.024$ \\
2455107.88 & $B$ & 15.626 (030) & $-0.027$ \\
2455113.79 & $B$ & 16.015 (030) & $-0.039$ \\
2455116.88 & $B$ & 16.188 (033) & $-0.025$ \\

2455065.33 & $V$ & 16.546 (052) &   0.031 \\
2455066.58 & $V$ & 15.827 (042) &   0.027 \\
2455068.58 & $V$ & 14.926 (031) &   0.017 \\
2455070.50 & $V$ & 14.411 (028) &   0.026 \\
2455073.00 & $V$ & 14.010 (024) &   0.017 \\
2455075.13 & $V$ & 13.816 (024) &   0.032 \\
2455076.63 & $V$ & 13.697 (025) &   0.038 \\
2455079.33 & $V$ & 13.561 (024) &   0.036 \\
2455080.63 & $V$ & 13.568 (024) &   0.034 \\
2455084.42 & $V$ & 13.602 (024) &   0.033 \\
2455086.75 & $V$ & 13.654 (024) &   0.044 \\
2455089.50 & $V$ & 13.748 (025) &   0.047 \\
2455098.96 & $V$ & 14.273 (028) &   0.043 \\
2455105.96 & $V$ & 14.515 (029) &   0.037 \\
2455107.88 & $V$ & 14.631 (030) &   0.037 \\
2455113.79 & $V$ & 14.925 (031) &   0.041 \\
2455116.88 & $V$ & 15.199 (034) &   0.036

\enddata

\tablecomments{Photometry given in magnitudes with 1$\sigma$
uncertainties (in mmags) presented in parentheses.  The noted
$S$-corrections have been applied to the photometry.}

\end{deluxetable}

To correct the KAIT and {\it Swift} photometry for small systematic
differences due to different instrumental response functions, we apply
$S$-corrections to put our optical photometry on the standard system
following the method of \citet{Stritzinger02}.  This procedure is done
in addition to standard color-corrections which are appropriate for
stellar SEDs, but can be inaccurate for the SED of a SN.  The
instrumental response function includes the effects of the filter
transmission, quantum efficiency of the detector, mirror reflectivity,
and the atmospheric transmission in the case of ground based
telescopes.  The instrumental response functions for our KAIT
photometry correspond to KAIT4 in \citet{Ganeshalingam10}.  We
calculate synthetic magnitudes for the SN using the Bessel
transmission curves of \citet{Stritzinger05} and the instrumental
response functions using our optical spectral series for SN~2009ig.
In instances where our spectra of SN~2009ig do not cover the $B$-band
(e.g., around maximum light), we instead use spectra from the
\citet{Hsiao07} spectral series at the appropriate phase warped using
a third-order spline to match the colors of the KAIT natural system
photometry.  The difference between the Bessel synthetic magnitude and
the color-corrected instrumental synthetic magnitude is the
$S$-correction \citep[See][for more details on computing
$S$-corrections]{Stritzinger02, Wang09:05cf}.  We fit splines to the
KAIT $B$ and $V$ data to estimate the KAIT magnitudes at the time of
the {\it Swift} observations in these bands.  The residuals of KAIT
minus {\it Swift} photometry in the $B$ ($V$) band have a mean and
standard deviation of $-0.01$~mag ($-0.03$~mag) and 0.07~mag
(0.06~mag), respectively.

\subsection{Optical Spectroscopy}

We obtained low- and medium-resolution optical spectra of SN~2009ig
with the Kast double spectrograph \citep{Miller93} on the Shane 3~m
telescope at Lick Observatory, the Low Resolution Spectrograph
\citep[LRS;][]{Hill98} on the Hobby-Eberly Telescope (HET) in queue
mode \citep{Shetrone07}, the Blue Channel spectrograph
\citep{Schmidt89} on the 6.5~m MMT telescope, the Low Resolution
Imaging Spectrometer \citep[LRIS;][]{Oke95} on the 10~m Keck~I
telescope, and the DEep Imaging Multi-Object Spectrograph
\citep[DEIMOS;][]{Faber03} on the 10~m Keck~II telescope.  Spectra
were typically made at low airmass and at the parallactic angle
\citep{Filippenko82}.  A log of our observations is given in
Table~\ref{t:spec}.

\begin{deluxetable*}{rllrl}
\tabletypesize{\scriptsize}
\tablewidth{0pt}
\tablecaption{Log of Optical Spectral Observations\label{t:spec}}
\tablehead{
\colhead{} &
\colhead{} &
\colhead{Telescope /} &
\colhead{Exposure} &
\colhead{} \\
\colhead{Phase\tablenotemark{a}} &
\colhead{UT Date} &
\colhead{Instrument} &
\colhead{(s)} &
\colhead{Observer\tablenotemark{b}}}

\startdata

$-14.2$ & 2009 Aug.\ 22.508 & Lick/Kast        & 1800                          & MC \\
$-14.1$ & 2009 Aug.\ 22.628 & Keck/LRIS        & 300                           & AS, HS \\
$-14.1$ & 2009 Aug.\ 22.635 & Keck/DEIMOS      & $2 \times 120$, $2 \times 60$ & ET, RB, RG \\
$-13.3$ & 2009 Aug.\ 23.428 & HET/LRS          & 900                           & SO \\
$-13.1$ & 2009 Aug.\ 23.623 & Keck/DEIMOS      & $4 \times 300$                & ET, RB, RG \\
$-12.2$ & 2009 Aug.\ 24.518 & Lick/Kast        & 900                           & VB \\
$-12.1$ & 2009 Aug.\ 24.638 & Keck/DEIMOS      & $3 \times 300$                & ET, RB, RG \\
$-11.2$ & 2009 Aug.\ 25.522 & Lick/Kast        & 900                           & VB \\
$-11.1$ & 2009 Aug.\ 25.632 & Keck/DEIMOS      & 300                           & ET, JK, RB \\
$-10.3$ & 2009 Aug.\ 26.495 & MMT/Blue Channel & 180                           & GW \\
$-10.1$ & 2009 Aug.\ 26.637 & Keck/DEIMOS      & $3 \times 300$                & ET, JK, RB \\
 $-9.3$ & 2009 Aug.\ 27.490 & MMT/Blue Channel & 180                           & FV, LJ \\
 $-9.3$ & 2009 Aug.\ 27.499 & Lick/Kast        & 600                           & SBC, DP \\
 $-8.3$ & 2009 Aug.\ 28.493 & MMT/Blue Channel & 180                           & FV, LJ \\
 $-8.2$ & 2009 Aug.\ 28.530 & Lick/Kast        & 1500                          & AM, JS, MK \\
 $-7.4$ & 2009 Aug.\ 29.421 & HET/LRS          & 450                           & SR \\
 $-7.3$ & 2009 Aug.\ 29.495 & MMT/Blue Channel & 120                           & FV, LJ \\
 $-6.4$ & 2009 Aug.\ 30.407 & HET/LRS          & 450                           & SR \\
 $-6.3$ & 2009 Aug.\ 30.440 & MMT/Blue Channel & 120                           & FV, LJ \\
 $-5.4$ & 2009 Aug.\ 31.402 & HET/LRS          & 550                           & JC \\
 $-5.3$ & 2009 Aug.\ 31.480 & MMT/Blue Channel & 120                           & GW \\
 $-3.4$ & 2009 Sep.\ 02.402 & HET/LRS          & 450                           & JC \\
 $-2.3$ & 2009 Sep.\ 03.490 & HET/LRS          & 450                           & SO \\
 $-1.4$ & 2009 Sep.\ 04.472 & HET/LRS          & 450                           & SO \\
 $-0.4$ & 2009 Sep.\ 05.404 & HET/LRS          & 450                           & SO

\enddata

\tablenotetext{a}{Days since $B$ maximum, 2009 Sep.\ 6.0 (JD 2,455,080.5).}

\tablenotetext{b}{AM = A.\ Morton, AS = A.\ Stockton, SBC = S.\ B.\ Cenko,
DP = D.\ Poznanski, ET = E.\ Tollerud, FV = F.\ Vilas, GW = G.\
Williams, HS = H.-Y.\ Shih, JC = J.\ Caldwell, JK = J.\ Kalirai, JS =
J.\ Silverman, LJ = L.\ Jiang, MC = M.\ Childress, MK = M.\
Kandrashoff, RB = R.\ Beaton, RG = R.\ Guhathakurta, SO = S.\ Odewahn,
SR = S.\ Rostopchin, VB = V.\ Bennert}

\end{deluxetable*}

Standard CCD processing and spectrum extraction were accomplished with
IRAF.  The data were extracted using the optimal algorithm of
\citet{Horne86}.  Low-order polynomial fits to calibration-lamp
spectra were used to establish the wavelength scale, and small
adjustments derived from night-sky lines in the object frames were
applied.  The DEIMOS data were reduced using a modified version of the
DEEP2 data-reduction
pipeline\footnote{http://astro.berkeley.edu/~cooper/deep/spec2d/.},
which bias corrects, flattens, rectifies, and sky subtracts the data
\citep{Foley07}.  We employed our own IDL routines to flux calibrate
the data and remove telluric lines using the well-exposed continua of
the spectrophotometric standards \citep{Wade88, Matheson00:2500,
Foley03}.  Our optical spectra are presented in Figure~\ref{f:spec}.

\begin{figure}
\begin{center}
\epsscale{1.15}
\rotatebox{0}{
\plotone{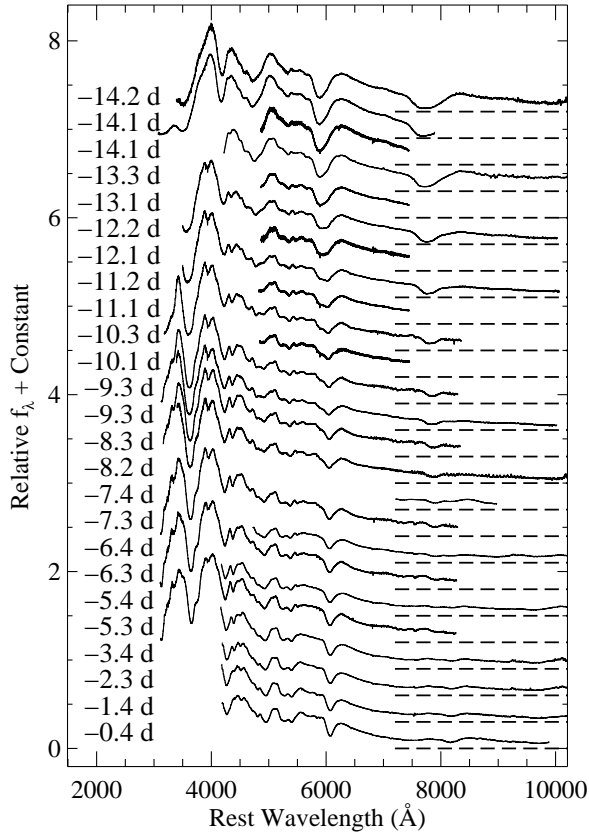}}
\caption{Optical spectra of SN~2009ig.  The spectra are denoted by
their phase relative to maximum brightness in the $B$ band.  The
zero-flux level is indicated for each spectrum by a dashed
line.}\label{f:spec}
\end{center}
\end{figure}

\subsection{Ultraviolet Spectroscopy}\label{ss:uvspec}

SN~2009ig was observed with the UV grism on {\it Swift}/UVOT for eight
pointings totaling 82608~s between 2009 August 23.7 and 2009 September
14.4.  The UVOT grism has spectral resolution $R = 150$, and a
wavelength accuracy of about 7~\AA.  The observations were obtained in
``clocked'' mode, which limits contamination by unrelated zero orders
at wavelengths longer than \about 3200~\AA.  The grism reduction
software had several updates compared to that used by
\citet{Bufano09}, including a wavelength equation that varied with
detector position, an extraction slit that followed the small spectral
curvature, and the use of a ninth-order polynomial to better follow
the rapidly changing background.  The gross spectrum was extracted
using a slit of width 13~pixels, and the background was extracted
within 11~pixels of the gross slit to minimize contamination by the
galaxy nucleus.  Due to an unfavorable roll angle, the background of
the first two spectra was contaminated by a bright F star, but there
was no obvious contamination of any other spectra.  To reduce the
contamination from the F star, background regions starting only 1
pixel from the SN were chosen for the first two spectra.  The
contamination is minimal, but there is still some slight
oversubtraction of flux for these spectra.  Details of our
observations are presented in Table~\ref{t:uvspec}.

\begin{deluxetable}{rlc}
\tabletypesize{\scriptsize}
\tablewidth{0pt}
\tablecaption{Log of {\it Swift}/UVOT Spectral Observations\label{t:uvspec}}
\tablehead{
\colhead{} &
\colhead{} &
\colhead{Exposure} \\
\colhead{Phase\tablenotemark{a}} &
\colhead{UT Date} &
\colhead{(s)\tablenotemark{b}}}

\startdata

$-13.0$ & 2009 Aug.\ 23.72 & 5742 \\
$-12.0$ & 2009 Aug.\ 24.72 & 6601 \\
$-11.1$ & 2009 Aug.\ 25.66 & 1491 \\
 $-9.2$ & 2009 Aug.\ 27.57 & 13803 \\
 $-4.2$ & 2009 Sep.\ 1.65  & 5764 \\
 $-2.1$ & 2009 Sep.\ 3.70  & 13318 \\
    1.5 & 2009 Sep.\ 7.34  & 17775 \\
    8.5 & 2009 Sep.\ 14.43 & 19259

\enddata

\tablenotetext{a}{Days since $B$ maximum, 2009 Sep.\ 6.0 (JD 2,455,080.5).}
\tablenotetext{b}{Exposure times are corrected for dead time.}

\end{deluxetable}


\section{Basic Observational Data}\label{s:basic}

SN~2009ig is a very well-observed SN.  Using our extensive photometry,
we provide basic parameters for it.  Fitting a polynomial through a
restricted region of each light curve, we determine the maximum-light
characteristics presented in Table~\ref{t:photdata}.

\begin{deluxetable*}{l@{\,\,}r@{\,\,}r@{\,\,}r@{\,\,}r@{\,\,}r@{\,\,}r@{\,\,}r@{\,\,}r}
\tabletypesize{\scriptsize}
\tablewidth{0pt}
\tablecaption{Photometric Information for SN~2009ig\label{t:photdata}}
\tablehead{
\colhead{Filter} &
\colhead{[$UVW2$]} &
\colhead{[$UVM2$]} &
\colhead{[$UVW1$]} &
\colhead{$U$} &
\colhead{$B$} &
\colhead{$V$} &
\colhead{$R$} &
\colhead{$I$}}

\startdata

JD of max.\ & \multirow{2}{*}{$78.30 \pm 0.05$} & \multirow{2}{*}{$81.30 \pm 0.05$} & \multirow{2}{*}{$76.76 \pm 0.05$} & \multirow{2}{*}{$76.57 \pm 0.08$} & \multirow{2}{*}{$80.54 \pm 0.04$} & \multirow{2}{*}{$82.17 \pm 0.03$} & \multirow{2}{*}{$81.33 \pm 0.29$} & \multirow{2}{*}{$78.26 \pm 0.25$} \\
\,\,\,\,$-$2,455,000 \\
Mag at max.\            & $16.50 \pm 0.04$  & $17.31 \pm 0.14$  & $14.75 \pm 0.04$  & $13.46 \pm 0.01$  & $13.66 \pm 0.03$  & $13.52 \pm 0.02$  & $13.64 \pm 0.02$  & $13.88 \pm 0.03$ \\
Peak abs.\ mag          & $-16.30 \pm 0.40$ & $-15.55 \pm 0.42$ & $-18.02 \pm 0.40$ & $-19.32 \pm 0.40$ & $-19.08 \pm 0.40$ & $-19.19 \pm 0.40$ & $-19.05 \pm 0.40$ & $-18.78 \pm 0.40$ \\
$\Delta m_{15}$ (mag)   & $1.03 \pm 0.03$   & $1.49 \pm 0.10$   & $1.10 \pm 0.03$   & $0.95 \pm 0.01$   & $0.89 \pm 0.02$   & $0.58 \pm 0.01$   & $0.69 \pm 0.01$   & $0.63 \pm 0.02$
 
\enddata

\end{deluxetable*}

SN~2009ig peaked at $V = 13.52$~mag, making it one of the brightest
SNe~Ia of the last decade.  The decline rate, $\Delta m_{15} (B) =
0.89$~mag, indicates that it was a slightly slower decliner than a
nominal SN~Ia with $\Delta m_{15} = 1.1$ mag.  Using Milky Way
reddening values \citep{Schlegel98} and the assumed distance modulus,
but neglecting any potential host-galaxy reddening, we find that
SN~2009ig had a peak absolute magnitude of $M_{V} = -19.19 \pm
0.40$~mag.  Fitting the KAIT \bvri\!\!\!  light curves with MLCS2k2
\citep{Jha07}, we find $\mu = 32.96 \pm 0.02$~mag (assuming $H_{0} =
74$~km~s$^{-1}$~Mpc$^{-1}$), consistent with the Tully-Fisher
measurement of $\mu = 32.6 \pm 0.4$~mag \citep{Tully88}.

The MLCS2k2 fit also finds a host-galaxy extinction of $A_{V} = 0.01
\pm 0.01$~mag, consistent with no host-galaxy reddening.  Examination
of the optical spectra shows that there is somewhat strong
\ion{Na}{1}~D absorption (equivalent width 0.4~\AA) at the redshift 
of the SN.  However, this level of absorption is consistent with zero
host-galaxy reddening as determined from large samples of SNe~Ia
\citep{Blondin09, Folatelli10, Poznanski11}.


\section{Ultraviolet Spectroscopy}\label{s:uv}

Under our program to get {\it Swift} UV spectra of nearby SNe~Ia
(GI--5080130; PI Filippenko), we obtained eight low-resolution UV
spectra of SN~2009ig.  Six spectra were from before maximum
brightness, a 38\% increase in the published premaximum UV spectra of
SNe~Ia.  The first spectrum was taken $-13.0$~days before $B$ maximum,
making it the earliest UV spectrum ever obtained of a SN~Ia.  The SN
was spectroscopically followed by {\it Swift} until 8.5~days after
maximum brightness.  The earliest spectrum shows a very broad
absorption feature spanning \about 3000--4000~\AA; however, it
separates into two distinct features (\ion{Co}{2} and Ca H\&K) only a
day later.  The UV continuum (2500--3000~\AA) is high in the first
spectrum, declining over the next 2--4~days, at which point it is
relatively similar for all following epochs.  The background region of
the first spectrum was slightly contaminated by a nearby star (see
Section~\ref{ss:uvspec}), but this has the affect of reducing the
continuum.

\begin{figure}
\begin{center}
\epsscale{1.1}
\rotatebox{90}{
\plotone{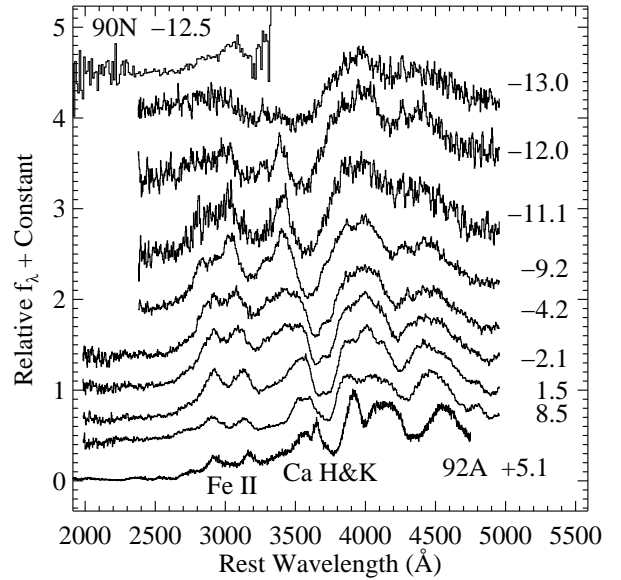}}
\caption{UV spectra of SN~2009ig.  The spectra are denoted by
their phase relative to maximum brightness in the $B$ band.  Also
plotted are the previously earliest spectrum of a SN~Ia, SN~1990N at
$t = -12.5$~days \citep{Leibundgut91}, and the earliest high-quality,
true UV spectrum of a SN~Ia, SN~1992A at $t = 5.1$~days
\citep{Kirshner93}.}\label{f:uvspec}
\end{center}
\end{figure}

The absorption feature at \about 3000~\AA, attributed to \ion{Fe}{2}
$\lambda$3250 \citep{Branch86}, is possibly present at $t \le -9.2$~days,
although it is quite weak.  Even at $t = -4.2$~days, it is not
particularly strong.  However, at $t = -2.1$~days, it has become rather
strong, and by $t = 1.5$~days, it is very strong, similar to that of
other SNe~Ia \citep{Foley08:uv}.

The \ion{Fe}{2} $\lambda$3250 feature is the strongest one in the UV.
Its velocity and pseudo-equivalent width (pEW) evolution have been
measured for several objects by \citet{Foley08:uv}, who found that the
velocity evolution is similar for all objects, but the strength of the
feature near maximum brightness correlates with light-curve shape.  We
present the velocity and pEW measurements for SN~2009ig relative to
the \citet{Foley08:uv} sample in Figures~\ref{f:vel_age} and
\ref{f:ew_age}, respectively.  SN~2009ig appears to follow the
previously established trends; its velocity is consistent with that of
other SNe~Ia, while its pEW is small, as expected given its relatively
broad light curve.

\begin{figure}
\begin{center}
\epsscale{0.9}
\rotatebox{90}{
\plotone{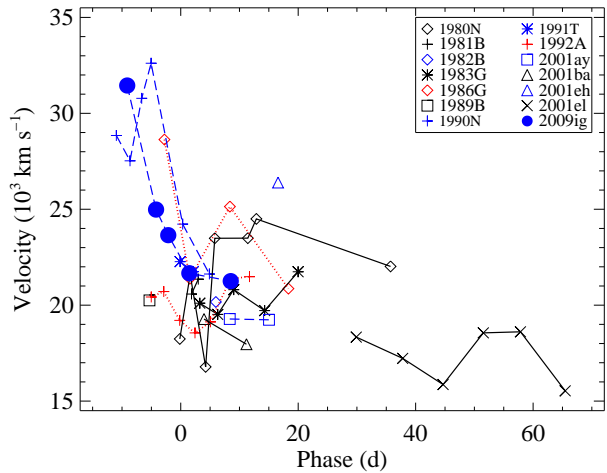}}
\caption{The blueshifted velocity of the minimum of the UV \ion{Fe}{2}
feature as a function of time for various SNe~Ia assuming a rest-frame
$gf$-weighted wavelength of 3250~\AA.  The normal, low, and
high-luminosity (as defined by \citealt{Foley08:uv}) SNe~Ia are shown
in black, red, and blue (with solid, dotted, and dashed lines
connecting points), respectively.}\label{f:vel_age}
\end{center}
\end{figure}

\begin{figure}
\begin{center}
\epsscale{0.9}
\rotatebox{90}{
\plotone{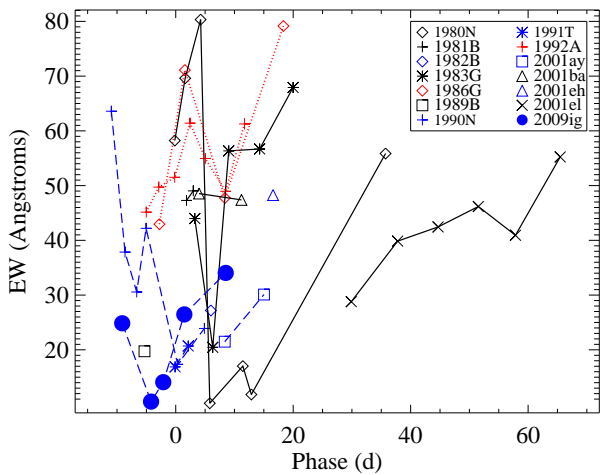}}
\caption{The pEW of the \ion{Fe}{2} feature as a function of time for
various SNe~Ia.  The normal, low, and high-luminosity (as defined by
\citealt{Foley08:uv}) SNe~Ia are shown in black, red, and blue
(with solid, dotted, and dashed lines connecting points),
respectively.}\label{f:ew_age}
\end{center}
\end{figure}

Overall, the UV spectral evolution of SN~2009ig is similar to that of
other normal SNe~Ia.  Near maximum brightness, we measure a UV ratio
\citep[$\mathcal{R}_{\rm UV} =
f_{\lambda}({\rm 2770~\AA})/f_{\lambda}({\rm
2900~\AA})$;][]{Foley08:uv} of 0.23, consistent with SN~2009ig having
a relatively broad light curve.  The shape of the SN~Ia UV SED is
determined by several factors related to the progenitor and
explosion.  High-quality UV spectroscopy can disentangle these effects
\citep{Sauer08}.  Unfortunately, the relatively low S/N data obtained
at $\lambda < 2500$~\AA\ precludes detailed modeling of the UV
spectrum similar to what was done by \citet{Sauer08}.


\section{Early-Time Photometry}\label{s:ep}

\subsection{Rise Time}

The first detection of SN~2009ig was 16.4~days before $B$ maximum, and
therefore the rise time for SN~2009ig was $\ge 16.4$~days.  The last
nondetection was 20.2~days before $B$ maximum.  However, extrapolating
the light curve in any reasonable manner, the last nondetection is not
sufficiently deep to be particularly constraining for the rise time.
Nonetheless, SN~2009ig is one of the earliest ever detected SNe~Ia.

Figure~\ref{f:rise} shows the light curves of SN~2009ig, shifted such
that the data represent the brightness below peak.  Assuming that a SN
is approximately a homologously expanding blackbody at early times,
the luminosity of the SN should increase as a function of $\tau^{2}$,
where $\tau$ is the time after explosion.  This form assumes that
there is negligible temperature evolution at early times, although
\citet{Arnett82} showed that this form should be independent of
temperature at sufficiently early times.  Given the strong color
evolution at early times (Section~\ref{ss:cc}), this model may not
apply to SN~2009ig.

Using a similar method to that of \citet{Riess99:risetime}, but
without stretching the light curve, we fit the $B$-band data with $t <
-10$~days to determine the rise time, $t_{r}$.  The data are best fit
with $t_{r} = 17.14 \pm 0.04$~days, where the uncertainty is only
statistical.  This fit is shown in Figure~\ref{f:rise}.  However,
there is a potential systematic difference in the measurement if the
SN is not well described by a homologously expanding blackbody.
Differences from this assumption should be larger at later times.  If
we examine only the first three KAIT data points (corresponding to
15.4, 14.5, and 12.4~days before $B$ maximum), the best-fit rise time
is $t_{r} = 17.11 \pm 0.07$~days, which is consistent with the other
value.  Nonetheless, we take the conservative approach of averaging
the two values and using the larger uncertainty for our final
measurement: $t_{r} = 17.13 \pm 0.07$~days.  Our first detection at $t
= -16.4$~days means that SN~2009ig was discovered \about 17~hours
after explosion.  Measured rise times for SN~2009ig in all optical
bands are consistently \about 17~days.  All photometric data indicate
that SN~2009ig was discovered $\lesssim$1~day after explosion.

\begin{figure}
\begin{center}
\epsscale{1.6}
\rotatebox{90}{
\plotone{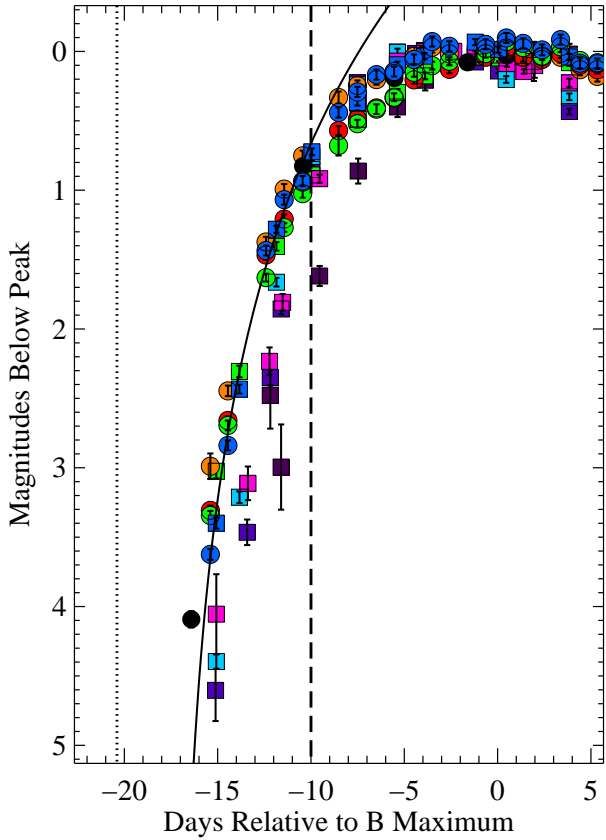}}
\caption{[$UVW2$][$UV\!M2$][$UVW1$]\ubvri\ (fuchsia, dark purple, navy,
cyan, blue, green, red, and orange, respectively; circles and squares
correspond to KAIT and {\it Swift}/UVOT data, respectively) and
unfiltered (black circles, with the label ``Unf'') light curves of
SN~2009ig.  The data have been shifted such that maximum brightness in
each band corresponds to zero magnitude.  The dotted line shows the
time of the last nondetection, while the dashed line corresponds to $t
= -10$~days. The solid curve represents the best-fit rise-time function
for the $B$ band with $t < -10$~days; the best-fit rise time is $t_{r} =
17.13$~days.}\label{f:rise}.
\end{center}
\end{figure}

\citet{Ganeshalingam11} examined a large sample of low-redshift SN~Ia
light curves, focusing on their rise time.  This sample included the
$B$ and $V$ KAIT light curves of SN~2009ig that are presented here.
However, the SN~2009ig light curves were not well fit by their
template light curve, and SN~2009ig was rejected from their final
analysis.

The rise time for SN~2009ig is relatively normal for a SN~Ia.
\citet{Hayden10:rise} determined that the average SDSS SN~Ia rise time
is $17.38 \pm 0.17$~days (with a standard deviation in the sample of
1.8~days) in the $B$ band.  \citet{Strovink07} found an average rise
time of $17.44 \pm 0.39$~days for a small sample of low-redshift
SNe~Ia with excellent premaximum light curves.  SN~2009ig has
high-velocity ejecta and a high velocity gradient for \ion{Si}{2}
$\lambda 6355$ (see \S~\ref{ss:es}), and \citet{Pignata08:02dj}
suggested that high velocity gradient SNe~Ia have shorter rise times
than low velocity gradient SNe.  \citet{Ganeshalingam11} found that
low-redshift high-velocity SNe~Ia had a rise time of $16.63 \pm
0.29$~days.  Observations of SN~2009ig support the trend that
high-velocity SNe~Ia tend to have shorter rise times in the $B$ band.

\citet{Hayden10:rise} and \citet{Ganeshalingam11} note that their
rise-time values are shorter than those found in previous studies
\citep[e.g.,][$t_{r} = 19.98 \pm 0.15$~days]{Riess99:risetime}, but the
differences are related to the adopted template light curve and a
single-stretch (rather than a two-stretch) method.  Our method does
not stretch the SN~2009ig light curves; we measure the true rise time
of SN~2009ig.

SN~2009ig declined in the $B$ band by 1.1~mag in 17.05~days;
therefore, the difference between the rise and fall time for SN~2009ig
is $t_{r} - t_{f} = 0.08$~days.  This value is consistent with the
majority of objects in the \citet{Hayden10:rise} sample, but is
slightly shorter than average. On the other hand,
\citet{Ganeshalingam11} found an average value of $t_{r} - t_{f} =
1.55 \pm 0.27$~days for high-velocity SNe~Ia.  The scatter is larger
than the quoted uncertainty in the mean, but there are few SNe in
their sample that have $t_{r} - t_{f} < 1$~day.

\subsection{Color Curves}\label{ss:cc}

Examining the optical color curves shown in Figure~\ref{f:cc}, we see
significant color evolution in $B-V$ for $t < -10$~days (this was also
noted by \citealt{Ganeshalingam11}).  From $t = -15.2$~days to $t =
-11.2$~days, $B-V$ decreased from 0.58~mag to 0.04~mag, a decline of
0.14~mag~day$^{-1}$.  The color changes in $V-R$ and $V-I$ are
relatively small at early times ($\lesssim 0.2$~mag).

\begin{figure}
\begin{center}
\epsscale{2.}
\rotatebox{90}{
\plotone{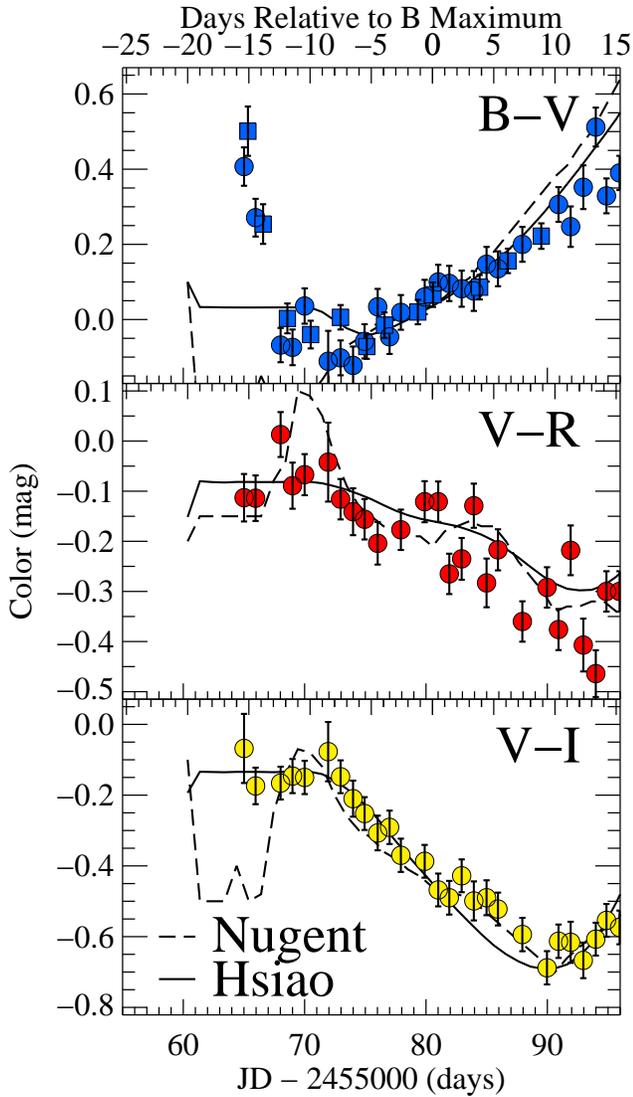}}
\caption{$B-V$, $V-R$, and $V-I$ color curves of SN~2009ig.  All
photometry has been corrected for the Milky Way reddening of $E(B-V) =
0.032$~mag \citep{Schlegel98}.  The solid and dashed lines represent
the color curves of the \citet{Hsiao07} and \citet{Nugent02} template
light curves, respectively, where each has been shifted to match
SN~2009ig at maximum brightness.}\label{f:cc}.
\end{center}
\end{figure}

The UV color curves (Figure~\ref{f:uvcc}) also show very fast color
evolution at early times.  $[UVW2]-V$, $[UV\!M2]-V$, and $[UVW1]-V$
all become redder by \about 1.5~mag from $t \approx -15$~days to $t
\approx -5$~days.  This is similar to the $U-V$ behavior, which is not
unexpected given the effective wavelengths of the filters for a SN~Ia
SED.  The color evolution is similar to the trends seen with a larger
sample \citep{Milne10}, but SN~2009ig extends these trends to much
earlier phases.

\begin{figure}
\begin{center}
\epsscale{2.}
\rotatebox{90}{
\plotone{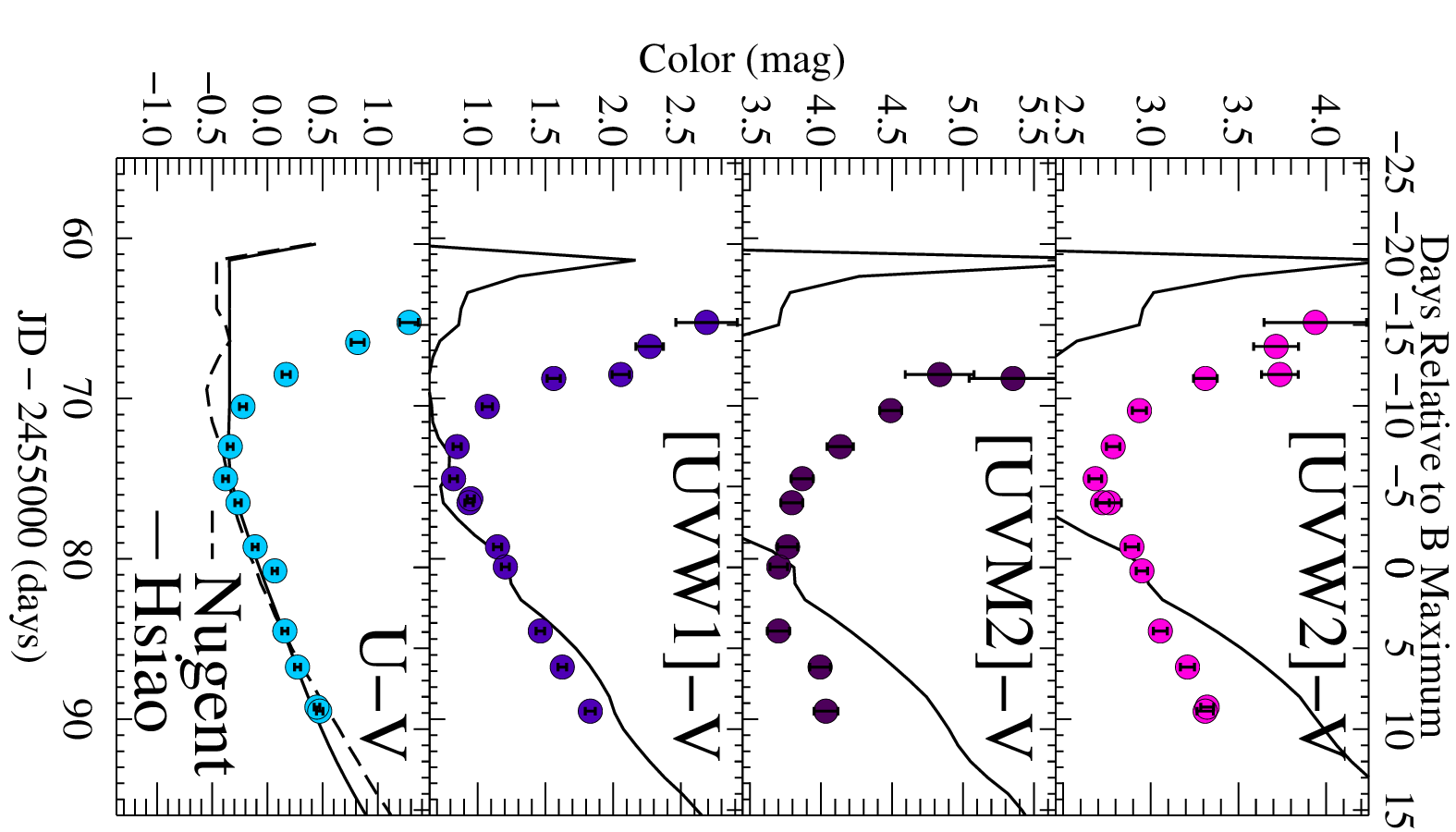}}
\caption{$[UVW2]-V$, $[UV\!M2]-V$, $[UVW1]-V$, and $U-V$ color curves
of SN~2009ig (all obtained with {\it Swift}).  All photometry has been
corrected for the Milky Way reddening of $E(B-V) = 0.032$~mag
\citep{Schlegel98} and the reddening corrections of \citet{Brown10}
for the {\it Swift} bands.  The solid and dashed lines represent the
color curves of the \citet{Hsiao07} and \citet{Nugent02} template
light curves, respectively, where each has been shifted to match
SN~2009ig at maximum brightness.}\label{f:uvcc}.
\end{center}
\end{figure}

Figures~\ref{f:cc} and \ref{f:uvcc} compare the color curves of
SN~2009ig to those of the commonly used SN~Ia template light curves of
\citet{Nugent02} and \citet{Hsiao07}.  After shifting to match the
colors of SN~2009ig at maximum brightness, the templates reproduce the
optical colors of SN~2009ig quite well for $t > -5$~days.  The
\citet{Nugent02} and \citet{Hsiao07} templates deviate from SN~2009ig
for $t \lesssim -6$ and $-9$~days, respectively, in $B-V$.  The $V-R$
colors appear to be more consistent out to the earliest observations
of SN~2009ig, but there may be a significant deviation at $t \approx
-10$~days.  The \citet{Nugent02} template matches the $V-I$ color of
SN~2009ig at even its earliest epochs, but the \citet{Hsiao07}
template deviates significantly at $t \lesssim -12$~days.

There are no \citet{Nugent02} template light curves for the UV bands,
but we are able to synthesize light curves in the {\it Swift} bands
from the \citet{Hsiao07} template spectra using the in-orbit filter
functions \citep{Poole08}.  The template matches SN~2009ig in $U-V$
and $[UVW1]-V$ after $t \approx -5$~days but is significantly bluer at
earlier times, similar to $B-V$.  The template light curves are
generally very poor in $[UVW2]-V$ and $[UV\!M2]-V$ at all times, but
are also bluer than the SN~2009ig data before maximum brightness (when
normalizing to the color at maximum brightness).  The long red tails
for the UV filters should not directly affect the synthesized light
curves if the filter definitions are correct.  However, the tails can
significantly affect the color curves depending on the reddening.

Since the \citet{Hsiao07} (and to a lesser extent at this point, the
\citealt{Nugent02}) template is used for comparing data to a standard
template and generating $K$-corrections for high-redshift SN~Ia light
curves, its fidelity is important.  Clearly, improvements can be made
at early times.  Nonetheless, the exact colors of the template are not
important for $K$-corrections (since the template SED is warped to match
the observed colors), but we show in Section~\ref{ss:es} that there
are significant differences between SN~2009ig and the \citet{Hsiao07}
template spectrum in the spectral features at early times.
Interestingly, \citet{Hayden10:rise} presents a K-corrected composite
SDSS SN~Ia $B-V$ color curve (their Figure~11) that has $B-V \approx 0
\pm 0.08$~mag for $t \le -10$~days, which is significantly different from
what is seen in SN~2009ig, but is consistent with the \citet{Hsiao07}
colors.  Although SN~2009ig may be a significant outlier, it is likely
that the $B-V$ color measured by \citet{Hayden10:rise} was
significantly affected by $K$-corrections which used the \citet{Hsiao07}
template.  Similar to the above cases, light-curve fitters, which also
use template light curves and SEDs, should consider these potential
problems.

\subsection{Emission from Interaction with a Companion Star}

There is a simple observational prediction for a SN~Ia that comes from
a single-degenerate progenitor system where the companion star fills
its Roche lobe: emission from the interaction of the SN ejecta with
the companion star that can dominate over the light produced from
radioactive decay at early times \citep{Kasen10:prog}.  The exact
characteristics of the emission, particularly the peak luminosity and
duration, depend on the progenitor system and viewing angle.  Relative
to the SN light curve, this emission should be brightest in the UV.
The effect strongly depends on viewing angle, where significant excess
emission is observed when the viewing angle is aligned such that the
companion star is between the WD and the observer, and essentially no
additional emission is generated for viewing angles 180$^{\circ}$ from
the WD-companion-observer configuration.  For all companions explored
by \citet{Kasen10:prog} (a 1~M$_{\sun}$ red giant, and 2 and
6~M$_{\sun}$ main-sequence stars --- all undergoing Roche-lobe
overflow so that the radius of the companion is approximately half the
separation distance), we would expect some excess emission in the $B$
band at $\tau \approx 2$~days if we observed the SN from \about
$10^{\circ}$ from the direction of the companion; the UV bands should
display an even stronger signal.

In Figures~\ref{f:lc} and \ref{f:rise}, we see no indication of early
excess emission in the optical bands.  However, there is some
indication of excess emission at early times in the UV.  Because of
their relatively low S/N, the [$UV\!M2$] data are not particularly
constraining.  However, the [$UVW2$] and [$UVW1$] data suggest some
excess emission.  We caution that because of the long red tail in
these bands, the apparent excess emission may be the result of optical
color evolution.

A strong indication that there is no detected interaction is that the
$B-V$ and UV colors of SN~2009ig at early times (our earliest
measurement is \about 0.7~day after explosion) are relatively red and
get bluer until $t \approx -10$~days.  This color evolution is not
expected for the \citet{Kasen10:prog} models.

To place further limits on excess emission at early times, the
\citet{Kasen10:prog} models are examined in more detail.  The
models produced are for the radioactively powered and interaction
powered components combined.  Since the \citet{Kasen10:prog} models do
not precisely reproduce the radioactively powered light curves, we
have decided to examine the difference between the light curves from
various viewing angles and the model light curve from a viewing angle
farthest from the companion (167$^{\circ}$).  There may be excess
emission from the interaction with the companion even in this model,
so the difference between the models can be considered a lower limit
on the expected emission for a given viewing angle.

Figure~\ref{f:kasen} shows the SN~2009ig light-curve data after
subtracting an expanding fireball model.  As explained above, this
should be a reasonable model for the early-time behavior of a SN~Ia.
For this comparison, we fit the data with $t < -10$~days relative to $B$
maximum, but ignored the first data point in each band.  The strongest
signal should be in the earliest data.  This results in a measurement
of the observed flux above that expected from modeling the additional
data.  If there is detectable excess flux in other observations, then
this should be apparent as systematic residuals to the expanding
fireball model.

\begin{figure}
\begin{center}
\epsscale{1.6}
\rotatebox{90}{
\plotone{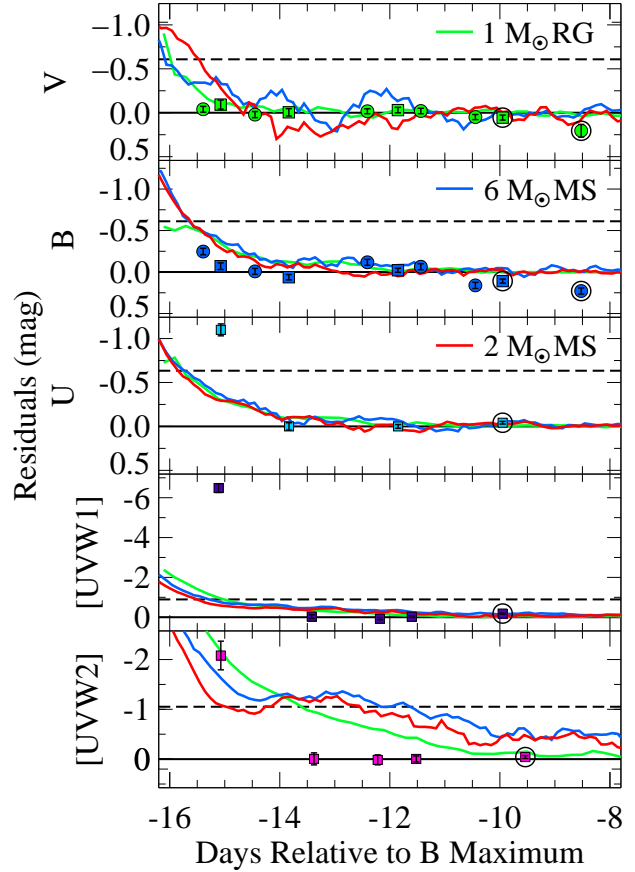}}
\caption{SN~2009ig [$UVW2][UVW1]U\!BV$ photometry (bottom to top panels) 
with an expanding fireball model subtracted.  The expanding fireball
model was fit to the data at $-15 < t < -10$~days relative to $B$
maximum, which excludes the first one or two data points.  The circled
points are at $t > -10$~days and are not included in the fit.  The
green, blue, and red curves correspond to excess emission in the
1~M$_{\sun}$ red-giant, 6~M$_{\sun}$ main-sequence, and 2~M$_{\sun}$
main-sequence companion \citet{Kasen10:prog} models at viewing angles
of 151$^{\circ}$, 109$^{\circ}$, and 68$^{\circ}$, respectively.
Specifically, these curves are an indication of the emission from the
interaction only.  These lines correspond to the largest viewing angle
(relative to the companion star) consistent with $B$-band observations
of SN~2009ig (i.e., are below the threshold at the time of the
earliest observation).  The dashed black line represents the threshold
above which a band has excess emission.}\label{f:kasen}.
\end{center}
\end{figure}

Because of their numerical limitations, the \citet{Kasen10:prog}
models have significant scatter in their flux from one epoch to
another, corresponding to \about 0.2~mag for each band.  This scatter
must be included in any potential detection.  We define our threshold
as three times the geometric mean of the model scatter and the
photometric uncertainty of the earliest data point in each band.  This
detection limit is marked as a dashed line in Figure~\ref{f:kasen}.

The first $V$-band observation is consistent with the expanding
fireball model.  The first $B$-band observation is slightly brighter
than expected from the expanding fireball model, but is not above our
threshold.  The first $U$-band measurement is significantly above our
threshold, but only two data points were used to constrain the model.
Both the [$UVW1$] and [$UVW2$] bands show significant excess in their
first observations relative to the expanding fireball model.  However,
the models were fit with only three data points in each of these
bands, so the fits may be biased.  Nonetheless, this trend of excess
flux in the UV bands is similar to what was seen in Figures~\ref{f:lc}
and \ref{f:rise}.  Furthermore, all bands bluer than $V$ have excess
flux in their first data point and the data slightly after $t = -10$~days
are consistent with the model, making an excess more likely.

Using the \citet{Kasen10:prog} model SEDs, we were able to construct
residual light curves in each of our bands.  Using the $B$ threshold,
we can place a limit on the possible viewing angle for the three
progenitors considered.  The 1~M$_{\sun}$ red-giant, 6~M$_{\sun}$
main-sequence, and 2~M$_{\sun}$ main-sequence models with viewing
angles (relative to the companion star) greater than 151$^{\circ}$,
109$^{\circ}$, and 68$^{\circ}$, respectively, are consistent with
this measurement (i.e., they have excess flux below the threshold at
the time of the earliest $B$-band observation).  Given the
distribution of possible viewing angles, these values corresponds to
$<$6\%, 34\%, and 69\% of the random viewing angles.  Despite being
formally inconsistent with a simple expanding fireball model, the
[$UVW2$] band is consistent with the \citet{Kasen10:prog} models for
the large angle considered above.  The excess emission in the $U$ and
[$UVW1$] bands is also in excess of these models/angles.  In fact, the
[$UVW1$] data are inconsistent with all models/angles investigated by
\citet{Kasen10:prog}.

The excess flux seen in the bluer bands appears to be real.
Supporting this view are that the difference from the expanding
fireball model is always an excess, data just after the phase range we
use to fit the model are consistent with the model, there appears to
be a visual excess in the bluest bands, and including the first data
point in a fireball fit significantly increases the $\chi^{2}$ per
degree of freedom. However, the disagreement with the models in the
various bands suggests that the simple \citet{Kasen10:prog} model is
not responsible for this excess.  Instead, considering that the colors
become bluer with time (the opposite of predictions) and the long tail
of the UV filters, we suggest that the excess seen in the UV light
curves is the result of optical (and perhaps UV) color evolution.  If
there is significant color evolution because of either changes to
spectral features or the temperature, then the expanding fireball
model may not be sufficient for modeling the early-time light curve.

Using samples of \about 100 SNe~Ia from SDSS and SNLS, respectively,
\citet{Hayden10:bump} and \citet{Bianco11} produced K-corrected
composite rest-frame $B$-band light curves.  The ensemble of light
curves did not show any deviation from a standard rise, indicating
that red-giant progenitor systems must be a small fraction (if there
is any contribution at all) of the SN~Ia progenitor systems.
\citet{Ganeshalingam11} examined 61 well-sampled low-redshift SN~Ia
light curves, finding no signs of interaction under simple
assumptions; however, relaxing these assumptions, they found large
systematic trends in the data, precluding any definitive conclusions.
Unlike previous work that focused on the statistical treatment of
large samples, we have concentrated on a single well-observed SN,
which can reduce some potential systematic biases.  Nevertheless, the
SN~2009ig data provide additional concern when making simple
assumptions about the early-time light curves of SNe~Ia.


\section{Optical Spectroscopy}\label{s:os}

\subsection{Early-Time Spectroscopy}\label{ss:es}

In Figure~\ref{f:early_spec}, we show a detailed comparison of a
subset of our earliest spectra.  In the first panel, we compare
spectra obtained $-14.2$ and $-13.3$~days relative to $B$ maximum.
Although these spectra are separated by only 1~day, there are some
differences, particularly in the shape of the \ion{Ca}{2}
near-infrared (NIR) triplet.  The second panel adds the spectrum at
$-12.2$~days relative to $B$ maximum.  This spectrum is significantly
different from that taken only two days earlier.  In particular, the
\ion{Ca}{2} NIR triplet and the \ion{Si}{2} $\lambda 6355$ features
have very different shapes.  The \ion{Si}{2} $\lambda 6355$ feature
has a flatter bottom and is much weaker at $t = -12.2$~days.  The
velocity of these features (as well as others, including Ca H\&K) is
significantly lower in the later spectrum.  The continuum is also
different, with the later spectrum being bluer, similar to what is
seen in the $B-V$ color (Figure~\ref{f:cc}).  The red wing of Ca H\&K
in the $t = -14.2$~days spectrum is non-Gaussian, with a change in the
slope at \about 3800~\AA.  The \ion{Si}{2} $\lambda 4130$ feature is
not distinct in the earliest spectrum.  By $t = -12.2$~days, this feature
is clear.  We speculate that at earlier times, this feature is at
higher velocity and simply blends with Ca H\&K, causing the peculiar
red wing of Ca H\&K; however, it is possible that ionization effects
suppress this feature at early times.  The last panel of
Figure~\ref{f:early_spec} adds the $t = -11.2$~days spectrum.  This
spectrum is very similar to the $t = -12.2$~days spectrum in feature
strength and continuum shape, but there is a noticeable difference in
the velocities of the various features.

\begin{figure}
\begin{center}
\epsscale{1.15}
\rotatebox{0}{
\plotone{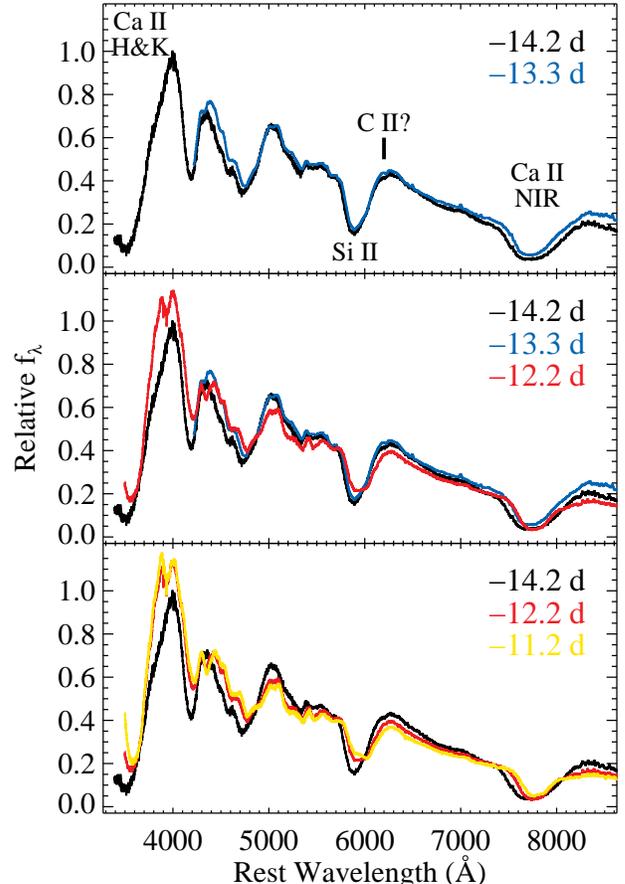}}
\caption{The $t = -14.2$ (black), $-13.3$ (blue), $-12.2$ (red), and
$-11.2$~days (yellow) spectra of SN~2009ig.  Features discussed in the
text are labeled.}\label{f:early_spec}
\end{center}
\end{figure}

\citet{Parrent11} analyzed two of the spectra presented here and
attributed a small depression at \about 6200~\AA\ to \ion{C}{2}
$\lambda 6580$.  Although we do not provide a detailed analysis of
this feature, at \about 6200~\AA, we note that it is present in all of
our first 6 spectra corresponding to $t \le -12.2$~days.  Those
spectra were obtained with 4 separate telescope/instrument
configurations, so we believe that it is a real feature.  It is also
present in our $t = -11.1$~days spectrum, but we do not have
definitive detections for our $t = -12.1$ or $-11.2$~days spectra,
though this is possibly the result of those spectra having a
relatively low S/N.  We do not detect this feature in any later
epochs, giving additional credibility to it being real.  If this
feature is \ion{C}{2} $\lambda 6580$, it would correspond to a
velocity of about $-17,000$~\kms, making it much lower velocity than
the velocity inferred from the minimum of \ion{Si}{2} at the same
epoch (about $-23,000$~\kms), but may be consistent with an
extrapolation of the lower-velocity \ion{Si}{2} component extrapolated
to the earliest epochs (see Section~\ref{ss:si}).  If \ion{C}{2} is
detected, this would be the first detection of it in a high-velocity
SN~Ia.

The spectra of SN~2009ig change dramatically until about 12~days before
maximum brightness.  After this time, the general spectral features
and shape stay relatively constant until after maximum brightness.
This epoch also approximately corresponds to the time when the SN
$B-V$ color evolution slows down (Section~\ref{ss:cc}).

In Section~\ref{ss:cc}, we showed that the early-time color evolution
of SN~2009ig was poorly matched by the color evolution of the
\citet{Hsiao07} template spectra.  However, that comparison is not apt
for $K$-corrections and some other applications since it neglected to
warp the template spectra to match the observed colors of SN~2009ig.
Without warping, the \citet{Hsiao07} template spectrum has different
broad-band colors than SN~2009ig.  We fit a spline to the ratio of the
broad-band integrated fluxes of the template and SN spectra.  The
template spectrum is divided by the spline to warp the template
spectrum to have the same \ubvri\!\!\! magnitudes of the combined
UV/optical SN~2009ig spectrum.  In the top panel of
Figure~\ref{f:kcor}, we show the $t = -13.0$~days {\it Swift} UV
spectrum, the $t = -14.2$~days optical spectrum, and the $t = -14$~days
\ubvri\!\!\! color-warped \citet{Hsiao07} template spectrum.  There
are several differences between the SN~2009ig and \citet{Hsiao07}
spectra.  In particular, the \citet{Hsiao07} spectrum has weaker,
lower-velocity features and significant deviations for
$\lambda < 4000$~\AA.

\begin{figure}
\begin{center}
\epsscale{1.8}
\rotatebox{90}{
\plotone{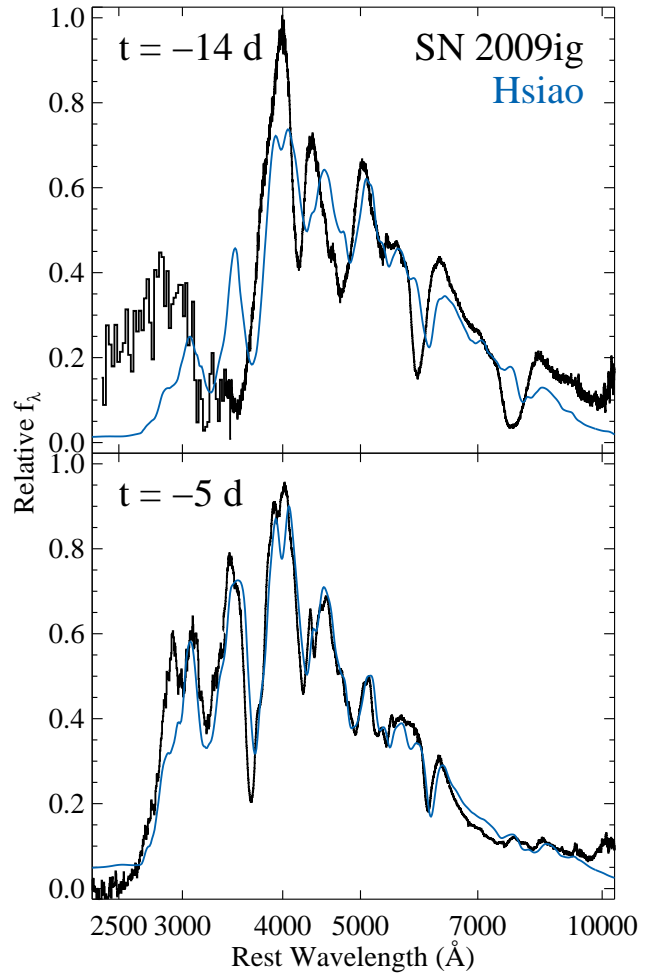}}
\caption{The $t = -13.0$~days {\it Swift} UV and $t = -14.2$~days optical
spectra (top panel) and the $t = -4.2$~days {\it Swift} UV and $t =
-5.8$~days optical spectra (bottom panel) of SN~2009ig (black curves).
The blue curves are the $t = -14$~days and $t = -5$~days \ubvri\!\!\!
color-warped \citet{Hsiao07} template spectra (top and bottom panels,
respectively).}\label{f:kcor}.
\end{center}
\end{figure}

The bottom panel of Figure~\ref{f:kcor} displays the $t = -4.2$~days {\it
Swift} UV spectrum, the $t = -5.8$~days optical spectra, and the $t =
-5$~days \ubvri\!\!\! color-warped \citet{Hsiao07} template spectrum.  At
this phase, the \citet{Hsiao07} template spectrum is a much better
match to SN~2009ig; there are still some differences in the UV,
particularly the feature at 2900~\AA\ and the continuum for $\lambda <
2600$~\AA, but the match is significantly better than for the $t =
-14$~days spectrum.  Clearly, additional early-time data are necessary to
improve the fidelity of template spectra and light curves.  Use of the
current template spectra will provide significantly discrepant
$K$-corrections, even after 5-band color warping.

\subsection{\ion{Si}{2} $\lambda 6355$}\label{ss:si}

While the overall shape of the spectrum of SN~2009ig shows most of its
evolution during the first few days after explosion, individual
spectral features continue to evolve for several more days.  An
example of this is the strong \ion{Si}{2} $\lambda 6355$ line.  Using
the method outlined by \citet[and references therein]{Foley11:vgrad},
we determined the wavelength of maximum absorption for this feature;
we present a characteristic velocity of the line in
Figure~\ref{f:sivel}.

\begin{figure}
\begin{center}
\epsscale{1.2}
\rotatebox{0}{
\plotone{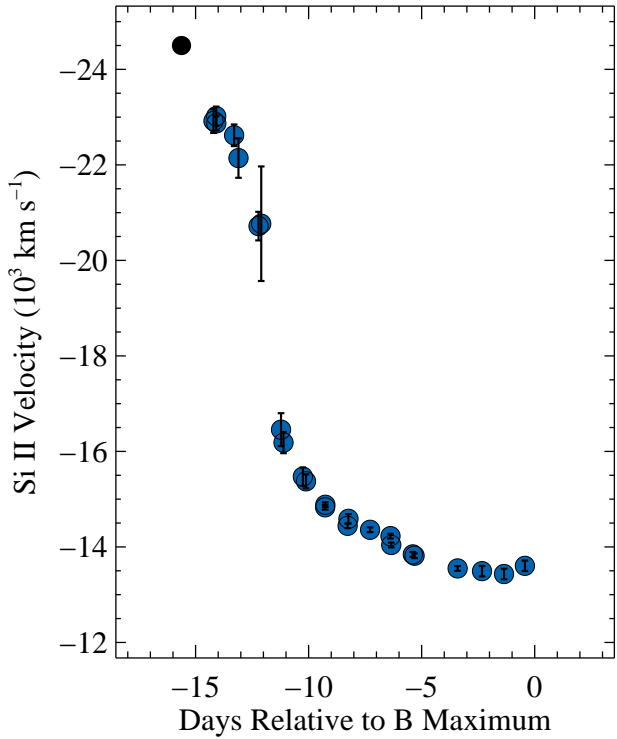}}
\caption{Velocity evolution of the \ion{Si}{2} $\lambda 6355$ feature
for SN~2009ig.  The blue points come from our spectra.  The black
point was reported by \citet{Navasardyan09}.}\label{f:sivel}.
\end{center}
\end{figure}

SN~2009ig displays the highest-velocity \ion{Si}{2} $\lambda 6355$
ever published, with the earliest spectra having $v = -23,000$~\kms\
at $t = -14.2$~days, significantly faster than the fastest measurements
of SN~2006X ($-20,700$~\kms\ on $t = -11.6$~days;
\citealt{Quimby06:06X}) and SN~2003W ($-21,000$~\kms\ on $t =
-11.2$~days; \citealt{Foley11:vgrad}).  \citet{Navasardyan09} reported
an even higher velocity of $-24,500$~\kms\ for SN~2009ig on $t =
-15.6$~days.  For our first spectrum, the blue wing of \ion{Si}{2}
$\lambda 6355$ extends to at least 5720~\AA, corresponding to $v =
-31,400$~\kms\ or $>$0.1$c$.

The velocity of the \ion{Si}{2} $\lambda 6355$ feature evolves very
quickly, decreasing by 5700~\kms\ in only 1.9~days.  After this dramatic
change, the velocity evolution appears to be linear in time, similar
to that of most SNe~Ia (\citealt{Foley11:vgrad}; Silverman et~al., in
prep.).  The velocity near maximum brightness is about $-13,500$~\kms,
which is faster than \about 85\% of all SNe~Ia with $1 \le \Delta
m_{15} (B) \le 1.5$~mag and faster than \about 85\% of all SNe~Ia with
$\Delta m_{15} (B) \le 1$~mag \citep{Foley11:vgrad}.

Further examination of the \ion{Si}{2} $\lambda 6355$ feature
demonstrates that the velocity evolution at early times is the result
of multiple components.  This is best displayed in Figure~\ref{f:si},
which shows the \ion{Si}{2} $\lambda 6355$ feature from spectra that
occur before, during, and after this velocity change.  At $t =
-13.1$~days, there is some indication of two components, with the blue
component dominating the profile.  The relative strength quickly
shifts; the components have similar strength at $t = -12.2$~days, but the
red component is dominant by $t = -11.2$~days.  This evolution continues
to the point where the blue component is barely visible at $t =
-9.3$~days.  Further examination and discussion of the velocity evolution
and line shapes of various features in the SN~2009ig spectra will be
given by Marion et~al.\ (in prep.).

\begin{figure}
\begin{center}
\epsscale{1.15}
\rotatebox{0}{
\plotone{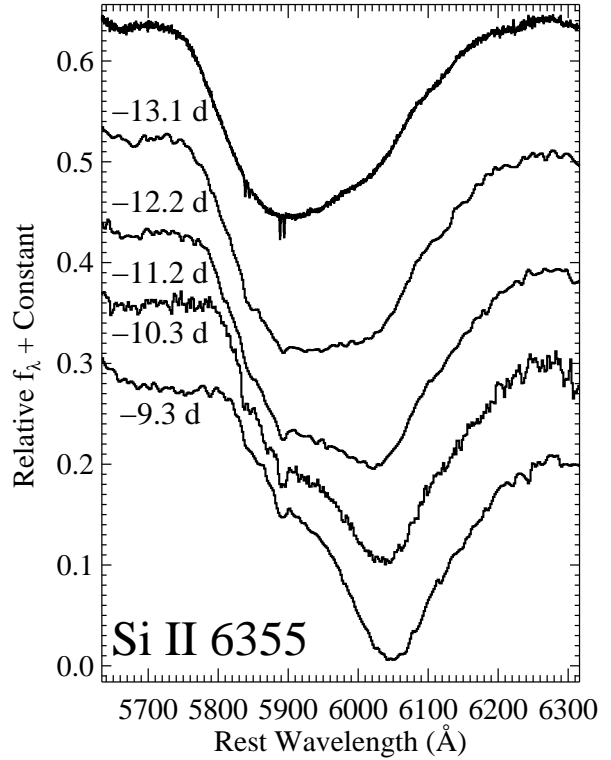}}
\caption{Spectra of SN~2009ig near the \ion{Si}{2} $\lambda 6355$
feature.  Phases are denoted for each spectrum.}\label{f:si}.
\end{center}
\end{figure}


\section{Discussion \& Conclusions}\label{s:disc}

SN~2009ig was discovered \about 17~hours after explosion, relatively
bright at peak ($V = 13.5$~mag), and well positioned for detailed
monitoring.  We obtained densely sampled UV/optical light curves, an
unparalleled UV spectral sequence, and an excellent, almost nightly
optical spectral sequence from discovery through maximum brightness.
Here we have focused on the early-time properties of this SN, as well
as on its unique UV spectral sequence.

SN~2009ig is a relatively normal SN~Ia, although it is a slightly slow
decliner ($\Delta m_{15} (B) = 0.89$~mag) and has relatively fast
ejecta at maximum brightness ($v^{0}_{\rm Si~II} \approx
-13,500$~\kms).  We find that the rise time in the $B$ band is
17.13~days, similar to that of other high-velocity SNe~Ia.  The UV
spectral features are similar to those of other SNe~Ia.  Our
observations, particularly the UV spectra and the early-time
photometry and spectroscopy, greatly expand the existing data at these
wavelengths and phases.

Despite being a relatively normal SN~Ia, early-time observations
deviate from template light curves and spectra: SN~2009ig is redder
than the templates.  Even after color warping an early-time template
spectrum to match a spectrum of SN~2009ig at a similar phase,
SN~2009ig has significantly broader, deeper, and higher-velocity
spectral features than the template.  The template also does a poor
job at matching both the spectral features and the continuum at
$\lambda < 4000$~\AA.  Distance estimates that rely on $K$-corrections
and/or early-time data from templates could be biased, directly
impacting cosmological measurements.  Studies that rely on early-time
K-corrected data, such as measuring the rise time \citep{Aldering00,
Conley06, Hayden10:rise} and interaction with a binary companion
\citep{Hayden10:bump, Bianco11} of high-redshift SNe~Ia, may be biased
by these differences.  Low-redshift studies should be minimally
affected.  An examination of the early-time light curve of SN~2009ig
shows excess emission relative to a simple expanding fireball model ($L
\propto \tau^{2}$) in the UV bands.  Although this behavior is
expected if there is interaction between the SN and a binary
companion, other predictions of the model (in particular, colors
becoming redder) are not seen.  It is therefore unlikely that we
detect any interaction in the case of SN~2009ig.

SN~2009ig displays two velocity components in the \ion{Si}{2} $\lambda
6355$ feature, with the higher-velocity component indicating the
highest-velocity SN~Ia ejecta ever observed. The relative strength of
the components changes dramatically from $t = -13$~days to $t =
-9$~days, and one would not infer the presence of a high-velocity
feature from spectroscopy after $t = -9$~days.  Since there are few
available spectra of SNe~Ia at such early epochs, high-velocity
\ion{Si}{2} $\lambda 6355$ may be ubiquitous or at least very common,
like high-velocity \ion{Ca}{2} features \citep{Mazzali05}.

In the age of giant synoptic surveys including that of the Large
Synoptic Survey Telescope, SNe similar to SN~2009ig will still be
rare.  Detailed studies of SNe that are as nearby, bright, well
positioned, and found as soon after explosion as SN~2009ig have the
potential to provide as much or more information for constraining
progenitor and explosion models than large samples of more distant
SNe.

\begin{acknowledgments} 

{\it Facilities:}
\facility{HET(LRS), Lick:KAIT, Lick:Shane(Kast), Keck:I(LRIS),
Keck:II(DEIMOS), MMT(Blue Channel), Swift(UVOT)}

R.J.F.\ is supported by a Clay Fellowship.  We thank the anonymous
referee for informed comments and suggestions.  We thank D.\ Kasen for
useful discussions.  We are grateful to the staffs at the Lick, Keck,
HET, and MMT Observatories for their dedicated services.  J.\ Bullock,
J.\ Caldwell, M.\ Kandrashoff, A.\ Morton, P.\ Nugent, S.\ Odewahn,
D.\ Poznanski, S.\ Rostopchin, H.-Y.\ Shih, F.\ Vilas, and G.\
Williams helped obtain some of the data presented here; we also thank
J.\ Lee and D.\ Tytler for attempting to obtain data.  {\it Swift}
spectroscopic observations were performed under program GI--5080130;
we are very grateful to N.\ Gehrels and the {\it Swift} team for
executing the program quickly.  Supernova research at Harvard is
supported by NSF grant AST--0907903.  A.V.F.'s supernova group at
U.C.\ Berkeley is supported by NASA/{\it Swift} grants NNX09AG54G and
NNX10AF52G, NSF grant AST--0908886, Gary and Cynthia Bengier, the
Richard and Rhoda Goldman Fund, and the TABASGO Foundation.  J.V.\
received support from Hungarian OTKA Grant K76816, NSF Grant
AST--0707769, and Texas Advanced Research Project grant ARP--0094.

Some of the data presented herein were obtained at the W.~M. Keck
Observatory, which is operated as a scientific partnership among the
California Institute of Technology, the University of California, and
NASA; the observatory was made possible by the generous financial
support of the W.~M. Keck Foundation.  The Hobby-Eberly Telescope
(HET) is a joint project of the University of Texas at Austin, the
Pennsylvania State University, Stanford University,
Ludwig-Maximilians-Universit\"{a}t M\"{u}nchen, and
Georg-August-Universit\"{a}t G\"{o}ttingen.  The HET is named in honor
of its principal benefactors, William P.\ Hobby and Robert E.\ Eberly.
We acknowledge the use of public data from the {\it Swift} data
archive.  KAIT was constructed and supported by donations from Sun
Microsystems, Inc., the Hewlett-Packard Company, AutoScope
Corporation, Lick Observatory, the NSF, the University of California,
the Sylvia \& Jim Katzman Foundation, and the TABASGO Foundation.
This research has made use of the NASA/IPAC Extragalactic Database
(NED), which is operated by the Jet Propulsion Laboratory, California
Institute of Technology, under contract with NASA.

\end{acknowledgments}

\bibliographystyle{fapj}
\bibliography{../astro_refs}


\end{document}